\documentclass[a4paper,11pt]{article}
\pdfoutput=1
\usepackage[dvipsnames]{xcolor}
\usepackage[T1]{fontenc} 
\usepackage[toc,page]{appendix}
\usepackage{jcappub} 
\usepackage{circledsteps}
\usepackage{natbib}
\setcitestyle{square,comma,numbers,sort&compress}
\usepackage{enumerate}
\usepackage{mathrsfs}
\usepackage{tabularx,booktabs}
\usepackage[left = 1.0in, top=1.0in,right=1.0in,bottom=1.0in,a4paper]{geometry}
\usepackage[caption=false]{subfig}
\usepackage[compat=1.1.0]{tikz-feynman} 
\usepackage{cancel}
\def\mc#1{\mathcal#1}
\usepackage[section]{placeins}
\usepackage{cleveref}

\usepackage{simpler-wick}
\newcolumntype{Y}{>{\centering\arraybackslash}X}

\definecolor{darkgreen}{rgb}{0,0.5,0}

\definecolor{myRED}{rgb}{0.8, 0.25, 0.33}

\title{\huge Gravitational Wave Probes of Massive Gauge Bosons at the Cosmological Collider}

\author[a]{Xuce Niu,}
\author[a,b]{Moinul Hossain Rahat,}
\author[a]{Karthik Srinivasan,}
\author[a]{and \quad \quad \quad Wei Xue}

\affiliation[a]{Institute for Fundamental Theory, Department of Physics,
University of Florida,\\Gainesville, FL 32611, USA }
\affiliation[b]{School of Physics \& Astronomy, University of Southampton, Southampton SO17 1BJ, UK }
\emailAdd{xuce.niu@ufl.edu}
\emailAdd{M.H.Rahat@soton.ac.uk}
\emailAdd{karthik.srinivas@ufl.edu}
\emailAdd{weixue@ufl.edu}

\abstract{ 
We extend the reach of the ``cosmological collider'' for massive gauge boson production during inflation from the CMB scales to the interferometer scales. Considering a Chern-Simons coupling between the gauge bosons and the pseudoscalar inflaton, one of the transverse gauge modes is efficiently produced and its inverse decay leaves an imprint in the primordial scalar and tensor perturbations. We study the correlation functions of these perturbations and derive the updated constraints on the parameter space from CMB observables. We then extrapolate the tensor power spectrum to smaller scales consistently taking into account the impact of the gauge field on inflationary dynamics. Our results show that the presence of massive gauge fields during inflation can be detected from characteristic gravitational wave signals encompassing the whole range of current and planned interferometers.
}

\makeatletter
\gdef\@fpheader{}
\makeatother
\begin{document}
\maketitle
\flushbottom

\section{Introduction}

Inflationary universe \cite{Guth:1980zm,Lyth:1998xn, Kinney:2003xf, Baumann:2009ds} is characterized by the Hubble scale $H$, which can be as large as $10^{14}$ GeV at the end of inflation. Such a high energy environment is a natural testbed for ultraviolet-scale physics that leaves its imprint on primordial fluctuations. In particular, the presence of new particles during inflation can be investigated from the ``squeezed'' limit of the three-point correlation function of the curvature perturbations, where the mass and spin of the particle are manifest in the frequency and angular distribution of the oscillatory bispectrum. This idea has been dubbed as ``cosmological collider'' \cite{Chen:2009zp, Chen:2009we, Baumann:2011nk, Arkani-Hamed:2015bza, Chen:2016nrs, Lee:2016vti, Meerburg:2016zdz, Chen:2016uwp, Chen:2016hrz,An:2017hlx,Kumar:2017ecc, Chen:2018xck,Wu:2018lmx,Li:2019ves, Lu:2019tjj,Hook:2019zxa, Hook:2019vcn, Kumar:2019ebj, Wang:2019gbi,Wang:2020uic, Wang:2020ioa, Maru:2021ezc, Lu:2021gso, Wang:2021qez, Tong:2021wai, Cui:2021iie, Pinol:2021aun, Tong:2022cdz, Reece:2022soh, 
Jazayeri:2022kjy, Pimentel:2022fsc, Chen:2022vzh, Qin:2022lva, Maru:2022bhr}, in analogy with terrestrial particle accelerators producing and detecting massive particles.

The production of massive particles during inflation is typically suppressed by a Boltzmann-like factor $\exp{(-\pi m/H)}$, where $m$ is the particle's mass and $H$ is the Hubble rate, the characteristic scale during inflation. It leads to a suppression of the signals at the cosmological collider. An interesting opportunity to overcome the Boltzmann suppression arises in the case of gauge bosons. If the inflaton $\phi$ is an axion-like pseudoscalar with approximate shift symmetry \cite{Freese:1990rb}, it can couple to gauge bosons through the dimension-5 Chern-Simons coupling $\phi F \tilde{F}/\Lambda$, where $F$ is the field-strengh of the gauge field, $\tilde{F}$ is its dual, and $\Lambda$ is a new scale. In this case, the production of one of the helicities of the gauge boson is enhanced by the factor $\exp{(\pi \xi)}$ \cite{Anber:2006xt,Anber:2009ua, Cook:2011hg, Barnaby:2010vf, Barnaby:2011qe, Barnaby:2011vw, Meerburg:2012id, Anber:2012du, Linde:2012bt, Cheng:2015oqa, Garcia-Bellido:2016dkw, Domcke:2016bkh, Peloso:2016gqs, Domcke:2018eki, Cuissa:2018oiw}, where $\xi \equiv \dot{\phi}_0/(2\Lambda H)$ is the so-called chemical potential, and $\phi$ is the homogeneous background value of the inflaton field $\phi$. For $m \sim \mc O(H)$, an $\mc O(1)$ chemical potential can overcome the Boltzmann suppression and lead to efficient massive gauge mode production from the decay of the inflaton.
The Chern-Simons coupling giving rise to particle production is typically Planck-suppressed. Unfortunately, the possibility of observing oscillatory bispectrum signals is bleak at cosmic microwave background (CMB)  or large scale structure (LSS) scales for such Planck-suppressed couplings \cite{Arkani-Hamed:2015bza, Dalal:2007cu, Matarrese:2008nc, Slosar:2008hx, Baumann:2012bc}.\footnote{See Ref.~\cite{Loeb:2003ya} for measurability using 21 cm tomography.}

In this paper, we open up a new observability window at much smaller scales through gravitational wave signals, which are produced via tensor fluctuations sourced by massive gauge bosons during inflation. 
%
The gravitational wave signals produced by massless gauge bosons are studied in \cite{Cook:2011hg,Domcke:2016mbx}. Creation of gravitational waves from tensor perturbations is a generic prediction of even the simplest models of inflation. So far it has evaded detection at the CMB scales, yielding a stringent upper bound on the almost scale-invariant gravitational wave amplitude. The presence of massive gauge fields acts as a new source of gravitational waves. Intriguingly, the production mechanism of the gauge fields is such that the sourced gravitational wave spectrum remains unobservably flat near the CMB scales, but rises at smaller scales to reach the sensitivity of gravitational wave detectors encompassing the nanoHz to kiloHz frequency range. This is a direct consequence of how inflaton rolls; while modes observable at CMB scales leave the horizon when `slow-roll' conditions prevail, detectable modes at gravitational wave interferometers originate after the departure from slow-roll. Inflaton's rolling speed increases while the Hubble rate decreases near the end of inflation, leading to an $\mc O(1)$ change in the chemical potential $\xi$, which exponentially enhances the gravitational wave signals. As a concrete example, we determine the evolution of the inflaton speed and the Hubble rate in the context of the generalized Starobinsky model \cite{Starobinsky:1980te} of inflation, which is currently favored by cosmological data \cite{Planck:2018jri}.

The parameter space spanned by the gauge boson mass and chemical potential is subject to various constraints at the CMB scales. The scalar power spectrum is precisely measured by COBE normalization \cite{Bunn:1996py} and WMAP \cite{WMAP:2010qai}, and the tensor power spectrum is tightly constrained from tensor-to-scalar ratio \cite{Planck:2018jri}. Furthermore, the nature of the perturbations created by the gauge fields is non-Gaussian, and there are strict bounds on scalar non-Gaussianity. Accounting for these bounds from updated cosmological data, we determine the allowed parameter space from which observable gravitational wave signals emerge. 

\begin{figure}[!ht] 
    \centering
      \includegraphics[width=0.99\textwidth]{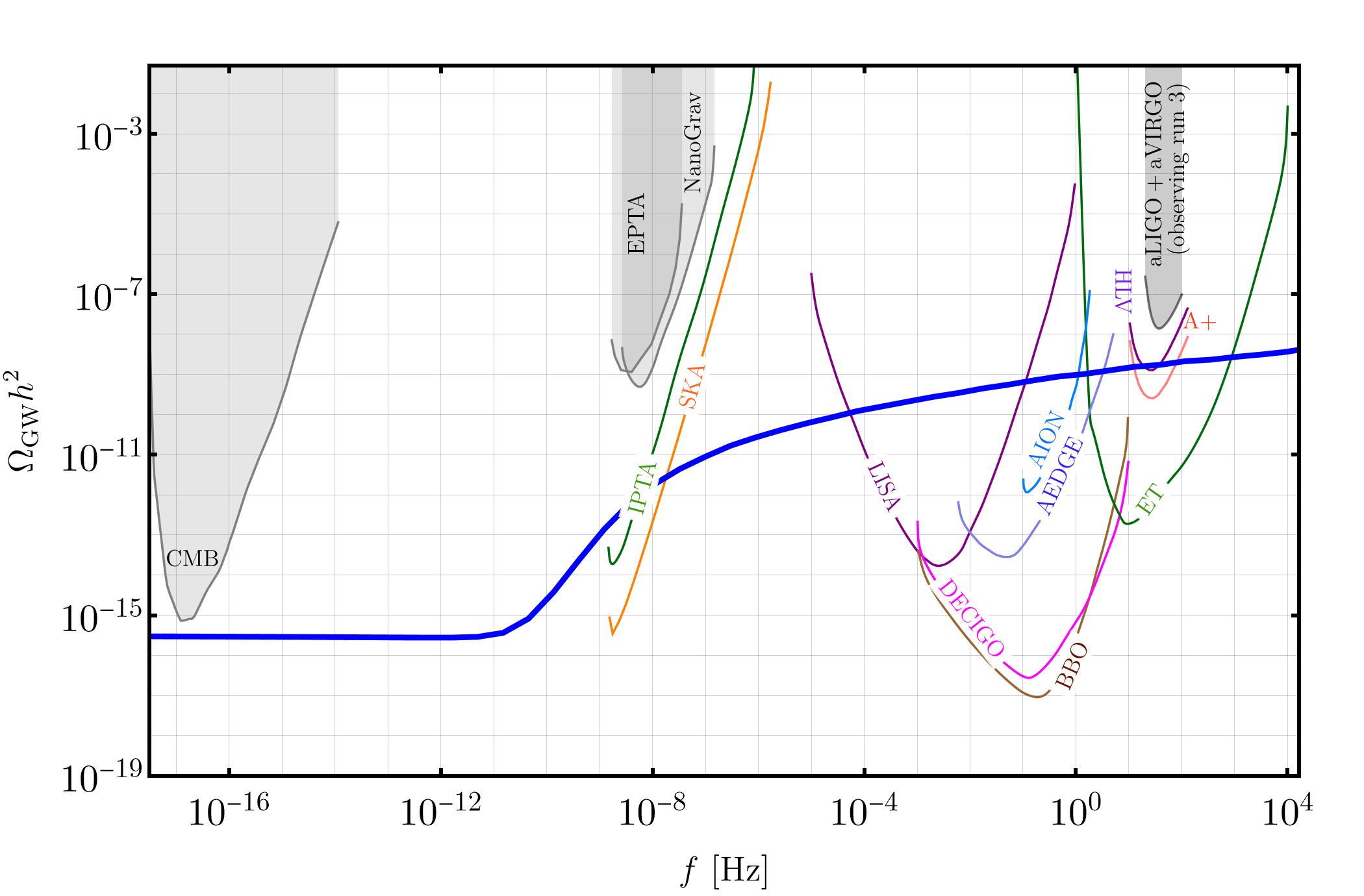}
    \label{fig:moneyplot}
    \caption{Gravitational wave amplitude for a benchmark point $m_A=1.5H, \xi_C = 2.45$. For comparison  we show the current upper bound (in gray) and future sensitivities (in color) of ongoing and proposed detectors, respectively. See text for details. 
    }
\end{figure}

An example of such a gravitational wave signal is shown in figure~\ref{fig:moneyplot} for the benchmark point $m_A = 1.5H$, $\xi = 2.45$ at the CMB scales corresponding to $f \sim 5 \times 10^{-17}$ Hz, in the context of the Starobinsky model of inflation.
The signal remains below the CMB upper bound, but starts to rise near $f \approx 10^{-10}$ Hz, and reaches the sensitivity of gravitational wave detectors from $1$ nHz to $1$ kHz while evading the upper bound from currently operational interferometers.

The paper is organized as follows. In section \ref{sec:2}, we discuss the mechanism of massive gauge field production from the Chern-Simons interaction. The contribution of the gauge field on scalar and tensor correlation functions is described in section~\ref{sec:4}. Backreaction effects of the produced gauge quanta on inflationary dynamics is discussed in section~\ref{sec:3}. In section~\ref{sec:5} we investigate the effect of various CMB constraints on the parameter space of the model. In section~\ref{sec:7} we present the gravitational wave signatures of massive gauge boson production during inflation. We discuss future directions and then conclude in section \ref{sec:8}.



\section{Massive Gauge Boson Production} \label{sec:2}
In this section we review the production mechanism of massive gauge fields from their coupling to the pseudoscalar inflaton during inflation. 

We start by considering a single field inflation theory where the pseudoscalar inflaton $\phi$ is coupled to a massive $U(1)$ gauge boson $A_\mu$ 
through the Chern-Simons interaction $\phi F \tilde{F}$. 
The coupling is suppressed by a new physics scale $\Lambda$, and the action takes 
the form
\begin{align}
    S = \int d^4x \sqrt{-g} \left[ -(\partial_\mu \phi) (\partial^\mu \phi)  
        - \frac{1}{4}F^{\mu\nu}F_{\mu\nu}+\frac{1}{2}m_A^2 A^\mu A_\mu - 
         \frac{1}{4\Lambda}\phi \Tilde{F}^{\mu\nu}F_{\mu\nu} \right], \label{action}
\end{align}
where $F_{\mu\nu} \equiv \partial_\mu A_\nu -\partial_\nu A_\mu$ is the field strength of the gauge field, 
and $\Tilde{F}^{\mu\nu} \equiv \frac{1}{2} \frac{\epsilon^{\mu\nu\alpha\beta}}{\sqrt{-g}}F_{\alpha\beta}$ is its dual, 
with $\epsilon^{0123}=+1$ is antisymmetric in any two indices. 
We assume a homogeneous, isotropic, expanding universe described by a  Friedmann-Robertson-Walker metric with a scale factor $a(t) = e^{Ht}$, where $H$ is the approximately constant Hubble rate during inflation. The metric can be expressed as 
\begin{align}
    ds^2 \equiv g_{\mu\nu}dx^\mu dx^\nu = dt^2 - a^2(t) dx_i dx^i = a^2(\tau) (d\tau^2 - dx_i dx^i),
\end{align}
where the cosmic time $t$ and conformal time $\tau$ are related by $d\tau = dt/a$. 
We introduce the usual notation $\dot{A} \equiv \partial_t A$ and $A' \equiv \partial_\tau A$. 
The Hubble rate and its conformal time counterpart are given by $H \equiv \dot{a}/a$ and $\mc H \equiv a'/a$, respectively.

From the action in \cref{action}, the equation of motion of the inflation can be written as
\begin{align}
    \ddot{\phi} + 3  H \dot{\phi} - \frac{1}{a^2(t)} \nabla^2\phi +  \frac{dV}{d\phi} = \frac{1}{\Lambda} \mathbf{E}\cdot \mathbf{B}, \label{EOMvarphi1}
\end{align}
while the Hubble rate can be obtained from the $00$ component of the Einstein equation
\begin{align}
    3  H^2M_{\rm Pl}^2 - \frac{1}{2}\dot{\phi}^2 - \frac{1}{2a^2(t)} (\mathbf{\nabla}\phi)^2 -   V = \frac{1}{2} \left[   \left(\mathbf{E}^2 + \mathbf{B}^2\right) + \frac{m_A^2}{a^2(t)} \mathbf{A}^2 \right].\label{EOMH}
\end{align}
Here we have introduced the physical electric and magnetic fields corresponding to the gauge field
\begin{align}
    \mathbf{E} = -\frac{1}{a^2} \mathbf{A'}, \qquad \mathbf{B} = \frac{1}{a^2} \boldsymbol{\nabla} \times \mathbf{A}. \label{electromagnetic}
\end{align}

Splitting the inflaton field into a homogeneous background part and a perturbation part 
\begin{align}
\phi(\tau, \mathbf{x}) \equiv \phi_0(t) + \delta \phi (t, \mathbf{x}),
\end{align}
we can express eqs.~\eqref{EOMvarphi1} and \eqref{EOMH} in terms of the background field taking the spatial mean of the source terms on the r.h.s.
\begin{gather}
    \ddot{\phi}_0 + 3 H\dot{\phi}_0+\frac{dV}{d\phi_0} = \frac{1}{\Lambda} \langle\mathbf{E}\cdot\mathbf{B}\rangle, \label{backr1}\\
    3 H^2 M_{\rm Pl}^2 - \frac{1}{2}\dot{\phi}_0^2 - V =  \frac{1}{2}\left\langle \mathbf{E}^2 + \mathbf{B}^2+\frac{m_A^2}{a^2} \mathbf{A}^2 \right\rangle. \label{backr2} 
\end{gather}

We now concentrate on the gauge field production by the rolling inflaton background $\phi(t)$. 
The field equation for the gauge field can be derived from the action \cref{action}. The four-divergence of the field equation yields the 
constraint $\partial_\mu (\sqrt{-g}A^\mu) = 0$, which leaves us with two transverse modes and a longitudinal mode.
The momentum space equation of motion for these can be expressed as
\begin{align}
    \partial_\tau^2 A_i + \left(k^2 + a^2(\tau) m_A^2\right)A_i - i \frac{\phi_0^\prime}{\Lambda} \epsilon_{ijk} k_j A_k + 2i \mc H k_i A_0 = 0. \label{EOMspatial}
\end{align}
From here
$A_0$ may be solved by the constraint $\partial_\mu (\sqrt{-g}A^\mu) = 0$.
The modes can be decoupled by decomposing the quantum field $\mathbf{A}$ as
\begin{align}
    \mathbf{A}(\tau, \mathbf{x}) = \sum_{\lambda = 0,\pm} \int \frac{d^3 k}{(2\pi)^{3}} \left[ \boldsymbol{\epsilon}_\lambda (\mathbf{k}) a_\lambda (\mathbf{k}) A_\lambda (\tau,k) e^{i \mathbf{k}\cdot \mathbf{x}} + \text{h.c.} \right], \label{gaugedecomp}
\end{align}
the longitudinal mode and the two transverse modes are denoted by $0$ and $\pm$, respectively. The creation/annihilation operators obey the commutation relation
\begin{align}\label{creationannihilation}
    \left[ a_{\lambda}(\mathbf{k}), a^\dagger_{\lambda'}(\mathbf{k'}) \right] = (2\pi)^3 \delta_{\lambda \lambda'} \delta^{(3)} (\mathbf{k} - \mathbf{k'}).
\end{align}
The polarization vectors have the following properties
\begin{align}
    \mathbf{k}\cdot \boldsymbol{\epsilon_{\pm}}(\mathbf{k}) = 0,\ \mathbf{k} \times \boldsymbol{\epsilon}_\lambda(\mathbf{k}) = - i \lambda k \boldsymbol{\epsilon}_\lambda(\mathbf{k}),\ \boldsymbol{\epsilon_{\pm}}^*(\mathbf{k}) = \boldsymbol{\epsilon_{\pm}}(-\mathbf{k}),\  \boldsymbol{\epsilon}_{\lambda}^*(\mathbf{k}) \cdot \boldsymbol{\epsilon}_{\lambda'}(\mathbf{k}) = \delta_{\lambda \lambda'}. \label{polarization}
\end{align} 
We can find the equation of motion of the transverse mode by applying \cref{gaugedecomp} in \cref{EOMspatial} and 
taking the dot product with $\boldsymbol{\epsilon}_\pm^*(\mathbf{k})$. This gets rid of the $A_0$ term in the equation since $\boldsymbol{\epsilon}_\pm^*(\mathbf{k})\cdot\mathbf{k}=0$. Hence $A_0$ does not affect the evolution of the transverse modes.
Further, considering the approximation of $a = -1/(H\tau)$ during inflation, the equation of motion for the transverse modes take the form
\begin{align}
    \partial_\tau^2{{A}_\pm}+ \left(k^2 + \frac{m_A^2}{H^2\tau^2} \pm \frac{2k\xi}{\tau}\right) {{A}_\pm}  &= 0, \label{eqpm2}
\end{align}
where we have defined the dimensionless chemical potential $\xi \equiv \dot{\phi}_0/(2\Lambda H)$. Strictly speaking, \cref{backr1,backr2,eqpm2} should be treated as a system of coupled equations for $\phi_0$, $H$ and $A_\pm$, which makes it very difficult to get an analytic solution for the mode functions. However, $\xi$ and $H$ change only marginally during inflation compared to the mode functions $A_\pm$, hence we can treat $\xi$ and $H$ as constants in \cref{eqpm2}.
Choosing the Bunch-Davies initial condition, \cref{eqpm2} then yields the following solution, up to a global phase, for the transverse modes
\begin{align}
    {A}_\pm &= \frac{1}{\sqrt{2k}} e^{\pm\pi \xi/2} W_{\mp i\xi, i\mu}(2ik\tau), \label{Whittaker}
\end{align}
where $W$ is the Whittaker W function,
and the parameter $\mu$ is defined as $\mu \equiv \sqrt{(m_A/H)^2 - 1/4}$. We will treat $\xi$ and $m_A$ as free parameters, which would be constrained from various observables to be discussed later.

Assuming $\dot{\phi}_0>0$ without loss of generality, we see that the $A_+$ ($A_-$) mode is enhanced (suppressed) by the chemical potential. 
The longitudinal mode is produced from purely gravitational interactions \cite{Graham:2015rva, Kolb:2020fwh, Ahmed:2020fhc}, but it is not affected by the chemical potential, as $\epsilon_{ijk}k_j A_k = 0$ for this mode in \cref{EOMspatial}. 
Since there is no enhancement for the longitudinal mode, it will not contribute much compared to the $+$ mode. 
In the following discussion we focus specifically on the `$+$' mode.

\begin{figure}[!ht]
    \centering
    \includegraphics[width=0.75\textwidth]{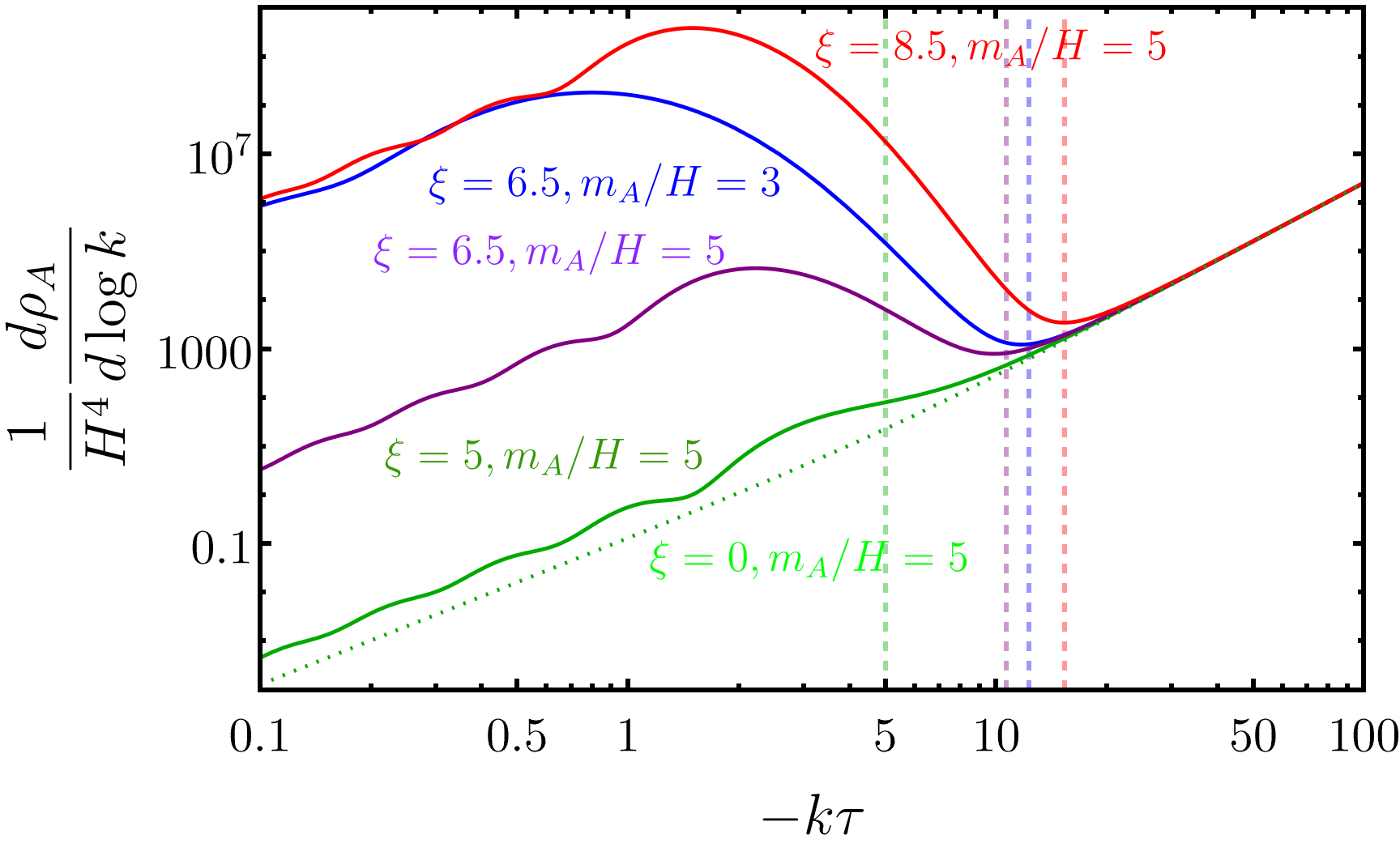}
    \caption{Evolution of the energy density of the gauge mode $A_+$ for a particular $k$ mode. Time flows from right to left. The vertical dashed lines at $-k\tau = \xi + \sqrt{\xi^2 - (m_A/H)^2}$ denote the boundary between the UV-divergent vacuum energy density (to the right of the line) and gauge mode energy density (to the left of the line). See text for details. 
    }
    \label{fig:Edensity}
\end{figure}

To illustrate the enhancement of the mode function with time, and how it depends on the parameters $\xi$ and $m_A$, we plot the dimensionless energy density per mode of the vector field with a particular comoving momentum $k$ in \cref{fig:Edensity} for five benchmark points. 
The average energy density of the gauge field can be expressed as 
\begin{align}
    \rho_A = \frac{1}{2}\left\langle \mathbf{E}^2 + \mathbf{B}^2 + \frac{m_A^2}{a^2} \mathbf{A}^2 \right\rangle,
\end{align}
from which we can define a dimensionless energy density per mode 
\begin{align}
    \frac{1}{H^4}\frac{d\rho_A}{d\log{k}} &= \frac{k^4\tau^4}{8\pi^2}e^{\pi\xi} \left[  
          \frac{1}{k^2}\left| \frac{dW}{d\tau} \right|^2   
       + \left(1+\frac{(m_A/H)^2}{k^2\tau^2}\right) |W|^2 
         \right], \label{energyden}
\end{align}
where $W \equiv W_{-i\xi,i\mu}(2ik\tau)$.
In \cref{fig:Edensity} we plot the r.h.s. of \cref{energyden} as a function of $-k\tau$, so that
it can be viewed as the evolution of a differential density spectrum with a fixed $k$, or the density spectrum at a given $\tau$. 
If we consider a fixed $k$, the plot shows
that time flows from right to left and  horizon crossing occurs at $-k\tau = p/H = 1$, where $p = k/a$ is the physical momentum. 
At the earliest times shown, when $-k\tau \gg 1$, gauge modes are deep inside horizon, and the energy density is dominated by the vacuum energy. 
The vacuum energy has UV divergences, which renormalize the cosmological constant and the Newton's gravitational constant \cite{Bunch:1978yq},
but it is unrelated to gauge field production. Since the particle production is dominant,
we set a   
hard cut-off at $-k\tau = \xi+\sqrt{\xi^2-(m_A/H)^2}$
when computing observable effects of the gauge modes. We choose this point because particle production happens mostly when the mode function experiences tachyonic instability, and from \cref{eqpm2}, this occurs for
\begin{align}
    -k \tau < \xi + \sqrt{\xi^2 - \left(\frac{m_A}{H}\right)^2}. \label{tachyonic}
\end{align} This cut-off is shown with dashed vertical lines in \cref{fig:Edensity}. 
The vacuum energy density drops at this point and then the gauge field energy density takes over, peaking near $-k\tau \simeq 1$, where copious particle production happens. After that, the energy density oscillates with a sharply decreasing envelope, representing the dilution by inflation. The frequency of oscillation is roughly proportional to the mass of the gauge field. The late time oscillation amplitude depends roughly on the difference between the mass and chemical potential, as seen from the red and blue curves. The amplitude of the peak is, however, dominated by the chemical potential, as seen from the red and the purple curves. Tachyonic instability vanishes when mass exceeds the chemical potential \cite{Meerburg:2012id}, as seen from the green curve. {The green dotted line shows the case for vanishing chemical potential, where even the late time oscillations flatten out.} The enhancement of the energy density showed in \cref{fig:Edensity} is due to the gauge field coupling with the inflaton through the chemical potential, and it is separated from the vacuum energy density. This leads to a clear distinction between the physical field amplification, and the standard
divergence associated with the empty vacuum state.

\section{Correlation Functions}\label{sec:4}
The production of massive gauge bosons and subsequent inverse decays during the inflationary era leaves their imprint both on the scalar and tensor perturbations. These effects can be understood from studying the power spectrum and non-Gaussianity through two- and three-point correlation functions. 
In this section we discuss the formulation for calculating these
correlation functions in term of the mode functions, and relegate the explicit details to the \cref{App:A}. The
in-in formalism \cite{Weinberg:2005vy} is employed to evaluate the correlation function. For the three-point correlation function, the real mode function approximation of $A_+$ 
applies to further simplify the formulas. 
The details and justification of real mode function approximation is given in ref.~\cite{Niu:2022fki}.



\subsection{Curvature Perturbation}

The equation of motion of the inflaton's perturbation can be obtained from subtracting \cref{backr1} from \cref{EOMvarphi1},
\begin{align}
    &\delta\ddot{ \phi} + 3 H \delta\dot{ \phi} - \left(\frac{1}{a^2}\nabla^2  - \frac{d^2 V}{d\phi^2}\right)\delta \phi   = \frac{1}{\Lambda} \left(\mathbf{E} \cdot \mathbf{B} - \langle \mathbf{E} \cdot \mathbf{B}\rangle\right). \label{EOMscalarperturb1}
\end{align}
In deriving this, we have not accounted for the fact that $\langle \mathbf{E}\cdot \mathbf{B} \rangle$ depends on $\dot{\phi}_0$; i.e. when replacing $\phi$ with $\phi_0+\delta \phi$, we must replace $\langle \mathbf{E}\cdot \mathbf{B} \rangle$ by $\langle \mathbf{E}\cdot \mathbf{B} \rangle$ $+$  $\delta \dot{\phi}\ \partial \langle \mathbf{E}\cdot \mathbf{B} \rangle / \partial \dot{\phi}_0$. This modifies the second term of \cref{EOMscalarperturb1} with a factor $\beta \equiv 1- 2\pi\xi {\langle \mathbf{E} \cdot \mathbf{B}\rangle}/{(3\Lambda H\dot{\phi}_0)}$
\begin{align}
     &\delta\ddot{ \phi} + 3\beta H \delta\dot{ \phi} - \left(\frac{1}{a^2}\nabla^2  - \frac{d^2 V}{d\phi^2}\right)\delta \phi   = \frac{1}{\Lambda} \left(\mathbf{E} \cdot \mathbf{B} - \langle \mathbf{E} \cdot \mathbf{B}\rangle\right). \label{EOMscalarperturb2}
\end{align}
Changing time variable to $\tau$, this becomes
\begin{align}
    \delta \phi'' + 2\mc H \delta\phi' - \frac{2\pi\xi a^2(\tau)}{\Lambda \phi_0^\prime} \langle \mathbf{E} \cdot \mathbf{B} \rangle \delta\phi' - \left( \nabla^2 - a^2 \frac{d^2V}{d\phi^2} \right) \delta \phi = \frac{a^2(\tau)}{\Lambda} \left( \mathbf{E} \cdot \mathbf{B} - \langle \mathbf{E} \cdot \mathbf{B} \rangle \right).
\end{align}
In the absence of the gauge fields, the classical vacuum solution to the inflaton perturbation can be expressed as
\begin{align}
    \delta \phi(\tau) = \frac{ H  }  { \sqrt{ 2 k^3} } ( 1 + i k \tau ) e^{ - i k \tau }
\end{align}
assuming the Bunch-Davies vacuum. 

In the presence of the gauge modes, we derive the correlation functions of the inflaton perturbation using the in-in formalism \cite{Weinberg:2005vy}. For some product of field operators $\mc Q(\tau)$, the correlation function in the in-in formalism can be expressed as 
\begin{equation}
   \begin{split}
   \langle {\cal Q} ( \tau ) \rangle
   &=
   \sum_{N= 0}^{\infty} i^N \int_{-\infty}^0 d\tau_N \int_{-\infty}^{\tau_N} d\tau_{N-1} \cdots  \int_{-\infty}^{\tau_2}  d\tau_1
      \, 
      \langle 
      [ \mathscr{H}_I ( \tau_1),  \cdots [\mathscr{H}_I (\tau_N), {\cal Q}_I ( \tau ) ] \cdots ] 
      \rangle 
   \, ,
   \end{split}
   \label{eq:inin2}
\end{equation}
where $\mathscr{H}_I(\tau)$ is the interaction part of the Hamiltonian, and $\mc Q_I(\tau)$ is the operator product in the interaction picture. 

For the Chern-Simons interaction in \cref{action}, the interaction part of the Hamiltonian can be written as $\mathscr{H}_I (\tau ) = - \int d^3x  \delta \phi J $, where $J$ is a source function given by
\begin{equation}
  J ( \tau,  {\bf x} ) = - \frac{1}{ 8 \Lambda}  \epsilon^{\mu \nu \rho \sigma } F_{\mu \nu } F_{\rho \sigma}.
\end{equation}
This form of the interaction Hamiltonian assumes that $J$ does not depend on $\delta \phi$ and $\delta \phi$ does not appear in the internal line.
In momentum space, the source function can be written as
\begin{equation}
   J_{\mathbf{k}}(\tau)  =  \frac{a^4(\tau)}{\Lambda} \int d^3x e^{-i \mathbf{k}\cdot \mathbf{x}} \mathbf{E} \cdot \mathbf{B} \label{Jk} \, .
\end{equation}

The curvature perturbation on uniform density hypersurfaces is defined as 
\begin{align}
\zeta (\tau, \mathbf{x}) \equiv -\frac{{H}}{\dot{\phi}_0}\ \delta \phi(\tau, \mathbf{x}).
\end{align}
The correlation functions of the curvature perturbation is calculated at $\tau_0 = 0$ after the end of inflation, and the classical value of the inflaton perturbation
becomes real, $\delta\phi ( 0  )  =  { H  }  / { \sqrt{ 2 k^3} } $.

\subsubsection*{Two-point correlation function}
From \cref{eq:inin2}, the two-point correlation function of the curvature perturbation due to the one-loop radiative correction from the gauge boson
can be expressed as
\begin{equation}
   \label{eq:2point}
   \begin{split}
    \langle  \zeta_{{\bf k}_1}  ( \tau_0 )   \zeta_{{\bf k}_2} ( \tau_0   ) \rangle_{(1)}
       &=  i^2 \left( - \frac{H}{\dot \phi } \right)^2
       \delta\phi_{k}^2(0 )
      \times 2 \, 
        \int_{-\infty}^0 d\tau_2  \int_{\infty}^{\tau_2} d\tau_1 
       \left(  \delta\phi (\tau _2)
        -   \delta\phi^* (\tau _2)  \right)
      \\
      & 
    \left[
      \delta\phi( \tau_1 )
       \langle
       J_{{\bf k}_1} ( \tau_1) 
       J_{{\bf k}_2} ( \tau_2 ) 
       \rangle
      -
      \delta\phi^*( \tau_1 )
       \langle
       J_{{\bf k}_2} ( \tau_2) 
       J_{{\bf k}_1} ( \tau_1 ) 
       \rangle
       \right].
   \end{split}
\end{equation}
Here the factor of $2$ accounts for the permutation of the momentum vectors $\bf{k}_1$ and $\bf{k}_2$,
and the subscript $(1)$ stand for the one-loop correction. The source function defined in \cref{Jk} can be expressed in terms of the the mode functions, using eqs.~\eqref{gaugedecomp} and \eqref{electromagnetic}, and considering only the $A_+$ mode,
\begin{align}
    J_{\mathbf{k}}(\tau) &= -\frac{1}{2\Lambda } \int \frac{d^3 q}{(2\pi)^{3}}  \left[ \boldsymbol{\epsilon}_{+}(\mathbf{q}) \cdot \boldsymbol{\epsilon}_{+}(\mathbf{k} - \mathbf{q}) \right] \nonumber \\
    &\times \left[\mc B_2(|\mathbf{k}-\mathbf{q}|,q;\tau)\ a_+(\mathbf{k}-\mathbf{q}) a_+^{}(\mathbf{q})  +  \mc B_2(|\mathbf{k}-\mathbf{q}|,\underline{q};\tau)\ a_+(\mathbf{k}-\mathbf{q}) a_+^\dagger(-\mathbf{q}) \right. \nonumber \\ 
    &+ \left. \mc B_2(\underline{|\mathbf{k}-\mathbf{q}|},{q};\tau)\ a_+^\dagger(\mathbf{q}-\mathbf{k}) a_+(\mathbf{q})  +   
    \mc B_2(\underline{|\mathbf{k}-\mathbf{q}|},\underline{q};\tau)\ a_+^\dagger(\mathbf{q}-\mathbf{k}) a_+^\dagger(-\mathbf{q})  \right].  \label{sourcefunc}
\end{align}
Here we have introduced a convenient notation  
\begin{subequations}
   \begin{align}
      {\cal B}_2 ( q_1 , q_2, \tau ) &\equiv q_1  A_+(\tau,q_1) A_+'(\tau,q_2) + q_2 A_+(\tau, q_2 )
         A_+'(\tau,q_1)  \, , 
      \\
      {\cal B}_2 ( \underline{q_1} , q_2, \tau ) &\equiv q_1  A_+^*(\tau,q_1) A_+'(\tau,q_2) + q_2 A_+(\tau, q_2 )
         A_+'^*(\tau,q_1)  \, , 
   \end{align}
\end{subequations}
so that an underline on the loop momentum on the l.h.s. denotes a complex-conjugation of the associated mode functions (and their derivatives) on the r.h.s. The source correlator is evaluated as
\begin{align}
     \langle J_{\mathbf{k_1}}(\tau_1)J_{\mathbf{k_2}}(\tau_2)\rangle &= \frac{(2\pi)^3\delta^{(3)} (\mathbf{k_1} + \mathbf{k_2})}{2\Lambda^2 } \int \frac{d^3 q_1}{(2\pi)^{3}} \left| \boldsymbol{\epsilon}_{+}(\mathbf{q_1}) \cdot \boldsymbol{\epsilon}_{+}(\mathbf{k_1} - \mathbf{q_1}) \right|^2 \nonumber \\
     &\times \mc B_2(|\mathbf{k_1}-\mathbf{q_1}|,q_1;\tau_1) \mc B_2(\underline{|\mathbf{k_1}-\mathbf{q_1}|},\underline{q_1};\tau_2).
\end{align}
Using this, the two-point correlation function in \cref{eq:2point} becomes
\begin{align}
\label{2pcfinin}
    &\langle  \zeta_{{\bf k}_1}  ( \tau_0 )   \zeta_{{\bf k}_2} ( \tau_0   ) \rangle_{(1)}^\prime
       =  -  \left( - \frac{H}{\dot \phi_0 } \right)^2 \frac{H^2} { 2 k^3} \frac{1}{\Lambda^2}
        \int_{-\infty}^0 d\tau_2  \int_{-\infty}^{\tau_2} d\tau_1 
      \int\frac{ d^3 q_1}{ ( 2\pi )^3} 
         | \boldsymbol{\epsilon}_+ ( {\bf q}_1 )  \cdot 
          \boldsymbol{\epsilon}_+ ( \mathbf{k_1} - \mathbf{q_1}  )  |^2 \nonumber
      \\
      &\times 
    \left[
      \delta\phi( \tau_1 )
       (  \delta\phi (\tau _2)
        -   \delta\phi^* (\tau _2)  )
      \mc B_2(|\mathbf{k_1}-\mathbf{q_1}|,q_1;\tau_1) \mc B_2(\underline{|\mathbf{k_1}-\mathbf{q_1}|},\underline{q_1};\tau_2) 
       + \rm{c.c.}
       \right],
\end{align}
where c.c. stands for the complex conjugate, and we have stripped off the $\delta$-function $(2\pi)^3\delta^{(3)}(\mathbf{k_1} + \mathbf{k_2})$ from the two-point correlator and set $|\mathbf{k}_1| = |\mathbf{k}_2| = k$.

Further simplification arises by exploiting the fact that the mode function in \cref{Whittaker} can have a global phase, which can be used to rotate the imaginary part away at a particular point. If the phase of the mode function remains fairly constant in a region, rephasing can make the mode function approximately real in that region. 
Intriguingly, this happens with the gauge boson mode function in the time domain relevant for particle production effects. We have verified that for $- k \tau \lesssim {\cal O}(1)$, which is the regime for copious gauge boson production, the Whittaker function has a nearly constant phase, which can be rephased away to make the mode function real. More details about this approximation can be found in ref.~\cite{Niu:2022fki}.
Under this approximation, ${\cal B}_2 (q_i,q_j, \tau ) = {\cal B}_2 ( \underline{q_i},q_j, \tau ) = {\cal B}_2 ( {q_i},\underline{q_j}, \tau ) =
{\cal B}_2 ( \underline{q_i},\underline{q_j}, \tau )  $, such that the time integrations in \cref{2pcfinin} can be decoupled
\begin{align}
\label{2pcfinin2}
    &\langle  \zeta_{{\bf k}_1}  ( \tau_0 )   \zeta_{{\bf k}_2} ( \tau_0   ) \rangle_{(1)}^\prime
       =  -  \left( \frac{H}{\dot \phi_0 } \right)^2 \frac{H^2} { 2 k^3} \frac{2^2}{\Lambda^2}
       \frac{1}{2}    
      \int\frac{ d^3 q_1}{ ( 2\pi )^3} 
         | \boldsymbol{\epsilon}_+ ( {\bf q}_1 )  \cdot 
          \boldsymbol{\epsilon}_+ ( \mathbf{k_1} - \mathbf{q_1}  )  |^2 
     \nonumber \\ 
      &\times 
    \int_{-\infty}^{0} d\tau_1\  \rm{Im}\
      \delta\phi( \tau_1 )\
      \mc B_2(|\mathbf{k_1}-\mathbf{q_1}|,q_1;\tau_1) 
     \int_{-\infty}^{0} d\tau_2\  \rm{Im}\
      \delta\phi( \tau_2 )\
      \mc B_2^*(|\mathbf{k_1}-\mathbf{q_1}|,q_1;\tau_2).
\end{align}
The factor $2^2$ comes from the two $2\ \rm{Im}\ \delta \phi(\tau_i)$, whereas the factor $1/2$ comes from changing the two-dimensional  integration region from a triangular to a rectangular region. 

\subsection*{Three-point correlation function}
The three-point correlation function can be derived using the in-in formalism as
\begin{eqnarray}
    &&
     \langle  \zeta_{{\bf k}_1}  ( \tau_0 )   \zeta_{{\bf k}_2}  ( \tau_0 )  \zeta_{{\bf k}_3}  ( \tau_0 ) 
      \rangle^\prime_{(1)}
      \nonumber
      \\
       &=&     
      i^3 \left( - \frac{H}{\dot \phi_0 } \right)^3   \frac{H^3} { ( 2  k_1 k_2 k_3   )^{3/2}} \frac{1}{\Lambda^3}
        \int_{-\infty}^0 d\tau_3 
        \int_{-\infty}^{\tau_3} d\tau_2  \int_{-\infty}^{\tau_2} d\tau_1 
      \nonumber
      \\
      && 
       \int\frac{ d^3 q_1}{ ( 2\pi )^3} 
          \boldsymbol{\epsilon}_+ ( {\bf q}_1 )  \cdot 
          \boldsymbol{\epsilon}_+ ( - {\bf q}_2  )   \, 
          \boldsymbol{\epsilon}_+ ( {\bf q}_2 )  \cdot 
          \boldsymbol{\epsilon}_+ ( - {\bf q}_3  )  \, 
          \boldsymbol{\epsilon}_+ ( {\bf q}_3 )  \cdot 
          \boldsymbol{\epsilon}_+ ( - {\bf q}_1  )  \,  
      \nonumber
      \\
      &&
    \bigg[
      \delta\phi_{k_1}( \tau_1 )
      \delta\phi_{k_2}( \tau_2 )
     (  \delta\phi_{k_3}( \tau_3 )
      -   \delta\phi_{k_3}^*( \tau_3 ) )
      {\cal B}_2 ( q_1 , q_2, \tau_1 ) 
      {\cal B}_2 ( \underline{q_2 } , q_3 ,  \tau_2 ) 
      {\cal B}_2 (  \underline{q_3} , \underline{q_1 } ,  \tau_3 ) 
      \nonumber
      \\
      &&
      - \delta\phi_{k_1}( \tau_1 )
      \delta\phi_{k_2}^*( \tau_2 )
     (  \delta\phi_{k_3}( \tau_3 )
      -   \delta\phi_{k_3}^*( \tau_3 ) )
      {\cal B}_2 ( q_1 , q_2, \tau_1 ) 
      {\cal B}_2 ( q_3  , \underline{q_1} ,  \tau_3 ) 
      {\cal B}_2 (  \underline{q_2} , \underline{q_3 } ,  \tau_2 ) 
      \nonumber
      \\
      &&
      +  {\rm 5 \, perms }
      - \rm{c.c.}
       \bigg],
\end{eqnarray}
where the loop momenta ${\bf q}_i$ flows into the vertex having the external field 
of momentum ${\bf k}_i$, and can be determined from conservation of momentum at each vertex
\begin{equation}
   {\bf q}_2  = {\bf q}_1  - {\bf k}_1 \, , 
      \quad
   {\bf q}_3  = {\bf q}_1  + {\bf k}_1 \, .
   \label{eq:q23}
\end{equation}
The $+ {\rm 5 \, perms}$ represents the permutation of $( {\bf k}_1 , {\bf k}_2, {\bf k}_3 )$ in the external lines for a fixed loop momentum configuration. Using the real model function approximation, we can derive a simpler expression for the three-point correlation function,
\begin{equation}
\label{3pcfinin2}
   \begin{split}
    &\langle  \zeta_{{\bf k}_1}  ( \tau_0 )   \zeta_{{\bf k}_2} ( \tau_0   ) \zeta_{{\bf k}_3}  ( \tau_0 ) \rangle'_{(1)}
       =   \left( \frac{H}{\dot \phi_0 } \right)^3 \frac{H^3} { (2k_1 k_2 k_3)^{3/2}} \frac{2^3}{\Lambda^3} \\
       &\times
      \int\frac{ d^3 q_1}{ ( 2\pi )^3} 
         \boldsymbol{\epsilon}_+ ( {\bf q}_1 )  \cdot 
          \boldsymbol{\epsilon}_+ ( - {\bf q}_2  )   \, 
          \boldsymbol{\epsilon}_+ ( {\bf q}_2 )  \cdot 
          \boldsymbol{\epsilon}_+ ( - {\bf q}_3  )  \, 
          \boldsymbol{\epsilon}_+ ( {\bf q}_3 )  \cdot 
          \boldsymbol{\epsilon}_+ ( - {\bf q}_1  ) 
      \\
      &\times 
    \int_{-\infty}^{0} d\tau_1\  \rm{Im}\
      \delta\phi( \tau_1 )\
      {\cal B}_2 ( q_1 , q_2, \tau_1 ) 
     \int_{-\infty}^{0} d\tau_2\  \rm{Im}\
      \delta\phi( \tau_2 )\
       {\cal B}_2 ( \underline{q_2 } , q_3 ,  \tau_2 ) \\
      &\times \int_{-\infty}^{0} d\tau_2\  \rm{Im}\
      \delta\phi( \tau_2 )\
      {\cal B}_2 (  \underline{q_3} , \underline{q_1 } ,  \tau_3 ).
   \end{split}
\end{equation}
Explicit expressions for the correlation functions can be found in \cref{App:A}.

\subsection*{Oscillatory bispectrum in the ``squeezed'' limit}
The two- and three-point correlation functions will be used to calculate the scalar power spectrum and the non-Gaussianity parameter $f_{\rm NL}$ at the CMB scale in  \cref{sec:5}. Here we would like to briefly comment on the ``cosmological collider'' signal --- the three-point correlation function in the ``squeezed'' limit, where one of the external momenta is much smaller compared to the other two, $k_3 \ll k_1 \approx k_2 = k$. The final expression in the dominant real mode function approximation is derived in \cref{App:A} and is given by
\begin{align}
    \langle\zeta_{\mathbf{k_1}}(\tau)\zeta_{\mathbf{k_2}}(\tau)\zeta_{\mathbf{k_3}}(\tau)\rangle'_{(1)} &= \frac{27}{256 \pi} {P_\zeta^{[\phi]}}^3 \xi^3 e^{3\pi\xi} \frac{1}{k^6}  \prod_{i=1}^{3} \int dx_i  \left(x_i \cos{x_i} - \sin{x_i}\right)\nonumber \\ &\times \mc W_1(y_3, 1;x_1) \
      \mc W_2(1,y_3;x_2)\   \mc W_3(y_3,y_3;x_3). \label{3PCFscalarCC}
\end{align}
Here $P_{\zeta}^{[\phi]}$ is the scalar power spectrum from vacuum fluctuations defined in \cref{scalarPinf}, and the $\mc W_i$ functions are given in eqs.~\eqref{W1}-\eqref{W3}. 

\begin{figure}
    \centering
    \includegraphics[width=0.65\textwidth]{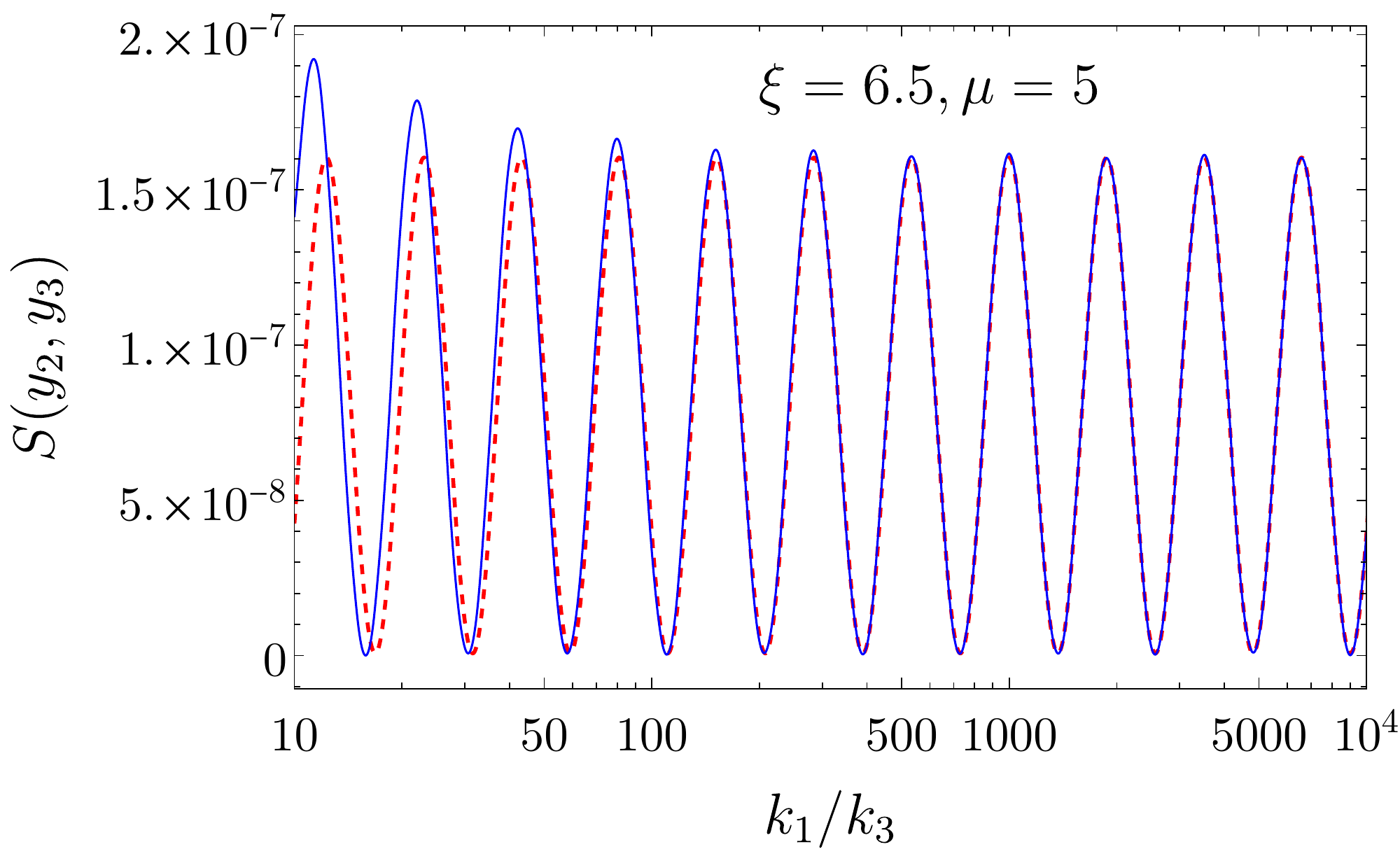}
    \caption{Oscillatory pattern of the scale-independent shape function of the scalar bispectrum in the squeezed limit for a benchmark point $\xi=6.5, \mu=5$. The solid blue line is the shape function calculated from \cref{shape} using \cref{3PCFscalarsq}. The red dashed line is a fitting function of the form $a + b \cos{\left[2\mu \log{(k_1/k_3)}+\vartheta\right]}$, showing that the oscillatory bispectrum's frequency is $2\mu = \sqrt{(2m_A/H)^2-1}$ with respect to $\log{(k_1/k_3)}$.}
    \label{fig:shapefunc}
\end{figure}

We can extract the overall momentum scale dependence by defining a `shape' function
\begin{align}
    S(y_2, y_3) &= {k^6\langle\zeta_{\mathbf{k_1}}(\tau_0)\zeta_{\mathbf{k_2}}(\tau_0)\zeta_{\mathbf{k_3}}(\tau_0)\rangle'_{(1)}}.  \label{shape}
\end{align}
In \cref{fig:shapefunc} we plot this `shape' function as a function of $ k_1/k_3 \equiv 1/y_3 $. It is an oscillatory function with frequency $2\mu$ and the envelope of its amplitude asymptotes to a constant for $k_1/k_3 \gg 1$. 

The behavior of the ``cosmological collider'' signal can be understood from using the late-time approximation \cite{NIST:DLMF} for the Whittaker functions involving $y_3 \ll 1$,
\begin{align}
    W(-2ix_1y_3) \approx (1-i)\sqrt{x_1 y_3} \left[ e^{\pi\mu/2} e^{i\mu \log{(2x_1 y_3)}} \frac{\Gamma(-2i\mu)}{\Gamma(\frac{1}{2} - i\mu +i\xi)} + (\mu \leftrightarrow -\mu)\right].
\end{align}
Using this, the $y_3$ dependence in \cref{3PCFscalarCC} is extracted to be of the form
\begin{align}
    (a + b y_3 + c y_3^2) e^{2i\mu \log{\left(y_3^{-1}\right)}},
\end{align}
where $a, b, c$ are constants. This is oscillatory in $\log{(y_3^{-1})}=\log{(k_1/k_3)}$ with a frequency $2\mu$. For smaller $y_3^{-1} = k_1/k_3$, the quadratic and linear terms in $y_3$ are dominant, whereas for $y_3^{-1} \gg 1$, the amplitude is dominated by the constant term.

\subsection{Tensor Perturbation}

Because of the exponential enhancement by the chemical potential, the gauge field can source large tensor modes in the primordial fluctuation \cite{Maleknejad:2011jw, Maleknejad:2011sq, Adshead:2012kp, Adshead:2013qp}. We use the scalar-vector-tensor decomposition of the perturbed metric and write it only in terms of the tensor perturbation $h_{ij}$
\begin{align}
    ds^2 = a^2(\tau) \left[ d\tau^2 - (\delta_{ij} + h_{ij})dx^i dx^j \right],
    \label{eq:hij_definition}
\end{align}
where $h_{ij}$ is transverse ($\partial_i h_{ij} = 0$) and traceless ($h_{ii} = 0$). The equation of motion of $h_{ij}$ is given by \cite{Weinberg:2008zzc}
\begin{align}
    h_{ij}''-\nabla^2 h_{ij} + 2 \mc H h_{ij}'=\frac{2}{M_{\rm Pl}^2} T_{ij}^{TT}, \label{eomh}
\end{align}
where $M_{\rm Pl} \simeq 2.4\times 10^{18}$ GeV is the reduced Planck mass, and $T_{ij}^{TT}$ is the transverse and traceless part of the stress-energy tensor. We decompose the tensor perturbation into two helicity modes
\begin{align}
    h_{ij}(\tau, \mathbf{p}) = \sum_{\lambda = \pm} \epsilon^\lambda_i(\mathbf{p})\epsilon^\lambda_j(\mathbf{p})\left(a_\lambda(\mathbf{p}) h_{p}^\lambda(\tau) + a^\dagger_\lambda(\mathbf{-p}) h_{p}^{\lambda*}(\tau)\right) \equiv
     \sum_{\lambda = \pm} \epsilon^\lambda_i(\mathbf{p})\epsilon^\lambda_j(\mathbf{p}) h^\lambda(\tau, {\bf p}) ,
\end{align}
where the creation/annihilation operators and polarization vectors obey \cref{creationannihilation,polarization}. The canonically normalized vacuum solution of \cref{eomh} in momentum space is given by
\begin{align}
    h_{{k}}^\lambda(\tau) = \frac{2H}{M_{\rm{Pl}} \sqrt{2k^3}}(1+i k\tau)e^{-ik\tau}.
\end{align}

The first order interaction term between the gravitational field and the vector boson field in the interaction Hamiltonian is given by ${a(\tau)}h_{ij}A_i'A_j'/{2}$. It can be expressed in terms of a source current
\begin{align} \label{intHtensor}
    \mathscr{H}_I =  \int \frac{d^3\mathbf{p}}{(2\pi)^3} h_{ij}(\tau, -\mathbf{p})J_{ij}(\tau,\mathbf{p}),
\end{align}
where the source current in momentum space is given by
\begin{align}\label{sourcetensor}
    J_{ij}(\tau, \mathbf{p}) = \frac{1}{2}\int\frac{d^3q}{(2\pi)^3} {\epsilon}_{+,i}(\mathbf{q})\epsilon_{+,j}(\mathbf{p}-\mathbf{q}) A_+'(\tau, \mathbf{q})A_+'(\tau, \mathbf{p}-\mathbf{q}).
\end{align}
We ignore the `$-$' mode of the vector field since it is not enhanced by the chemical potential, and drop the subscript `$+$' from now on. 

Using the in-in formalism, we can write the one-loop radiative correction to two-point correlation function of the tensor perturbations as
\begin{align}
    \left\langle h_{ij}(\tau_0,{\mathbf{k}_1})h_{ij}(\tau_0,{\mathbf{k}_2}) \right\rangle_{(1)} = -\int_{-\infty}^0 d\tau_2 \int_{-\infty}^{\tau_1} d\tau_1 \left 
    \langle\left[H_I(\tau_1),\left[H_I(\tau_2),h_{ij}(\tau_0,{\mathbf{k}_1})h_{ij}(\tau_0,{\mathbf{k}_2})\right]\right]\right \rangle,
\end{align}
where the correlation function is evaluated at $\tau_0 = 0$ at the end of inflation.
Plugging in \cref{intHtensor,sourcetensor} into this, and separating the two helicities, the two-point function becomes
\begin{align}
    \left\langle h^{\lambda}(\tau_0, \mathbf{{k}_1})h^{\lambda}(\tau_0, \mathbf{{k}_2})\right\rangle_{(1)}^\prime &= - \frac{2H^2}{M_{\rm Pl}^2}  \frac{1}{k^3} \int\frac{d^3\mathbf{p}}{(2\pi)^3}   \left|\boldsymbol{\epsilon}_{-\lambda}(\mathbf{k}_1)\cdot\boldsymbol{\epsilon}_+(\mathbf{p})\right|^2\ \left|\boldsymbol{\epsilon}_{-\lambda}(\mathbf{k}_1)\cdot\boldsymbol{\epsilon}_+(\mathbf{k}_1-\mathbf{p})\right|^2\nonumber\\ &\times \int_{-\infty}^0 d\tau_2 \int_{-\infty}^{\tau_2} d\tau_1\ A'(\tau_1, p)A'(\tau_1, {|\mathbf{k}_1-\mathbf{p}|})  A'^{*}  (\tau_2, p)A'^{*}(\tau_2, {|\mathbf{k}_1-\mathbf{p}|}) \nonumber\\ &\times h^{\lambda}_{{k}_1}(\tau_1)\left(h^{\lambda}_{{k}_1}(\tau_2)-h^{\lambda*}_{{k}_1}(\tau_2)\right)   + \rm{c.c.}\label{inin2pcftensor}
\end{align}
where we have stripped off the delta function $(2\pi)^3\delta^3(\mathbf{k}_1+\mathbf{k}_2)$.
Further simplification can be achieved using the real mode function approximation, as discussed in \cref{App:A}. This becomes particularly useful for calculating three- and higher point correlation functions. For our phenomenological study, tensor three-point correlation function is not of interest, as the current bounds from tensor non-Gaussianity are very relaxed. Nevertheless, for completeness, we include the final expression for the tensor three-point correlation function in the dominant real mode function approximation in \cref{App:A}.  

In our numerical study for tensor power spectrum and gravitational wave amplitude, we use the in-in result of \cref{inin2pcftensor} for greater accuracy. We have verified that the dominant real mode function approximation yields results in the same order of magnitude.

\section{Backreaction Effects} \label{sec:3}
So far we have ignored the time evolution of the chemical potential and the Hubble rate in our analysis. This is a reasonable assumption at least up to the CMB scale, as the slow-roll condition prevails, and the Hubble rate is nearly constant. However, modes that leave the horizon in later stages of inflation may be subject to strong backreaction effects from the inverse decay of the gauge field. Backreaction modifies the evolution of the Hubble rate and the rolling speed of the inflaton, thus also affecting the chemical potential. 




The effects of the vector field on the inflaton can be studied by taking the mean of the equation of motion of the inflaton and the Friedmann equation given by \cref{backr1,backr2}. These represent a pair of coupled equations for $H$ and $\dot{\phi}_0$ with respect to time, where the terms on the r.h.s. are source terms from the gauge field contribution. Substituting for $\mathbf{E}$ and $\mathbf{B}$ using \cref{electromagnetic}, we get
\begin{align}
    \langle \mathbf{E} \cdot \mathbf{B} \rangle &= - \frac{1}{4\pi^2 a^4} \int dk\ k^3 \frac{d}{d\tau} |A_+|^2 \equiv \frac{H^4}{8\pi^2}e^{\pi\xi}  I_1, \label{cond1}\\
    \frac{1}{2}\langle \mathbf{E}^2 + \mathbf{B}^2+\frac{m_A^2}{a^2} \mathbf{A}^2 \rangle &= \frac{1}{4\pi^2 a^4} \int dk\ k^2 \left( |{A_+}'|^2+(k^2+a^2 m_A^2) |A_+|^2 \right) \equiv \frac{H^4}{8\pi^2}e^{\pi\xi} I_2 \label{cond2},
\end{align}
where the integrals $I_1$ and $I_2$ are defined as 
 \begin{align}
     I_1 &\equiv \int_{0}^{x_{\rm max}}\ dx\ x^3 \frac{d}{dx}\left|W\right|^2, \\
 I_2 &\equiv \int_{0}^{x_{\rm max}}\ dx\ x^3 \left[ \left| \frac{dW}{dx}  \right|^2  + \left(1+\frac{(m_A/H)^2}{x^2}\right) \left| W \right|^2 \right],
 \end{align}
where $x\equiv -k\tau$, and $W \equiv W_{-i\xi, i\mu}(-2ix)$.
We cut off the integrals at $x= x_{\rm max} \equiv \xi+\sqrt{\xi^2-(m/H)^2}$ following the discussion below \cref{energyden}. %
Solving the coupled equations \eqref{backr1} and \eqref{backr2} gives the evolution of the inflaton speed and the Hubble rate with time.

We assume that the backreaction effects are negligible at CMB scales. This can be ensured by restricting ourselves to the parameter space where the source terms in \cref{backr1,backr2} are negligible compared to terms on the l.h.s. This implies
\begin{align}
    \pi\xi+\log{I_2}+2\log{\frac{H}{M_{\rm Pl}}}-5.47 &\ll 0, \label{BR2} \\
    \log{\xi}+\pi\xi+\log{|I_1|} + \log{P_\zeta^{[\phi]}} -1.1 &\ll 0. \label{BR1}
\end{align}
In the parameter space satisfying the above two constraints, inflationary dynamics is determined by the homogeneous solution of eqs.~\eqref{backr1} and \eqref{backr2}. We adopt this parameter space for observables at CMB scales. However, primordial perturbations responsible for observables at smaller scales leave the horizon later than the CMB modes when the backreaction of the produced gauge modes may become significant. 

To incorporate backreaction effects we need to evolve eqs.~\eqref{backr1} and  \eqref{backr2} simultaneously.
Typically, the source term in \cref{backr2} is negligible compared to the source term in \cref{backr1} and can be ignored. It is convenient to change variables from time to the efolding number $N$ left before the end of inflation, where $dN = -H dt$. In this convention $N$ decreases as we approach the end of inflation. Eqs. \eqref{backr1} and \eqref{backr2} can then be expressed as
\begin{gather}
    \frac{d^2\phi}{dN^2} + \frac{d\phi}{dN} \left( 3+\frac{d\ \log{H}}{dN} \right) + \frac{1}{H^2} \frac{dV}{d\phi} = \frac{1}{H^2} \frac{1}{\Lambda} \langle \mathbf{E} \cdot \mathbf{B} \rangle, \label{phiEOM} \\
    H^2 \approx V\ \left[3-\frac{1}{2}\left( \frac{d\phi}{dN} \right)^2\right]^{-1}. \label{HEOM}
\end{gather}

\begin{figure}[!ht]
    \centering
    \subfloat[ \label{fig:xivsN}]{
        \includegraphics[width=0.499\textwidth]{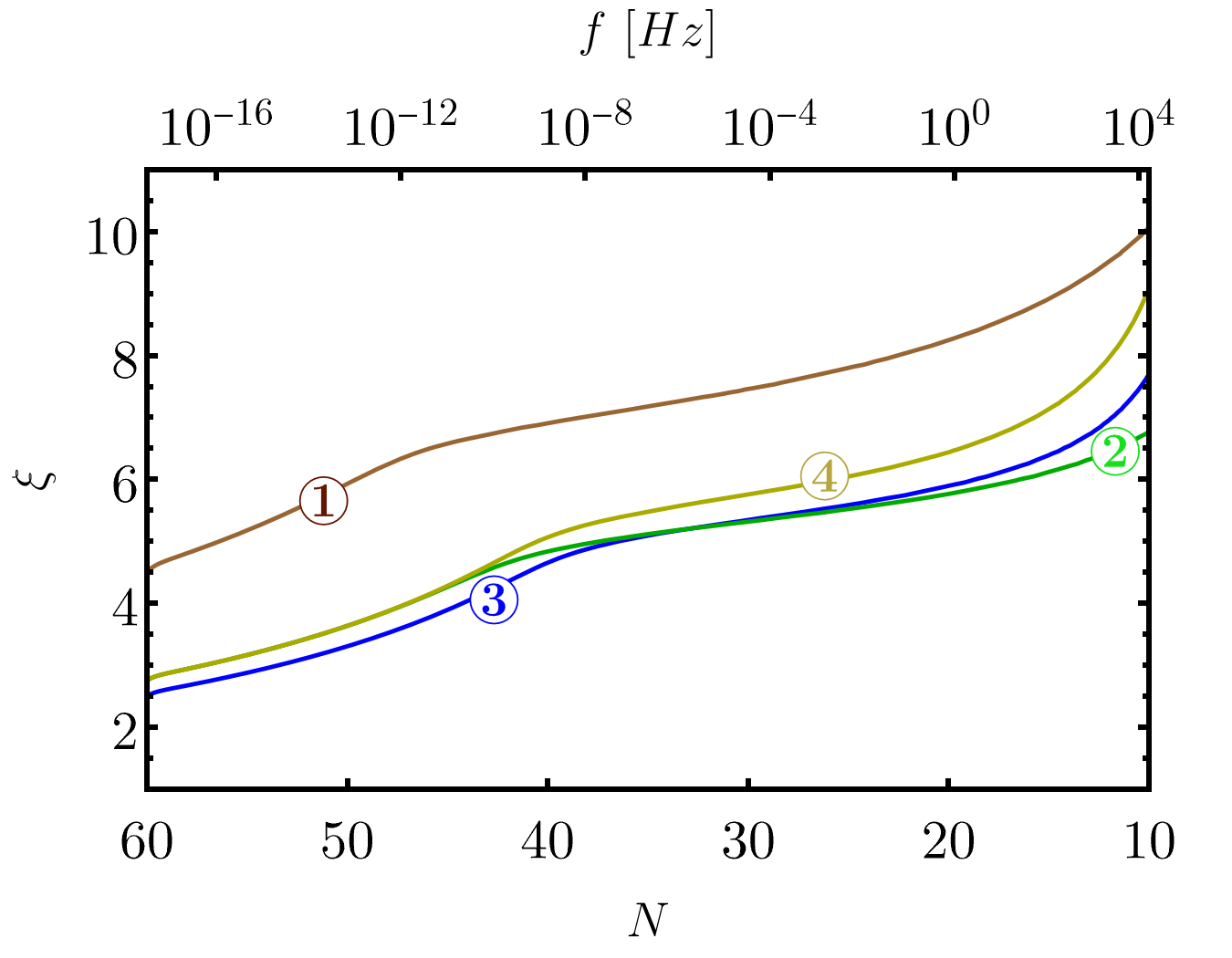}
    }
    \hspace{-0.6cm}
    \subfloat[ \label{fig:massvsN}]{
        \includegraphics[width=0.499\textwidth]{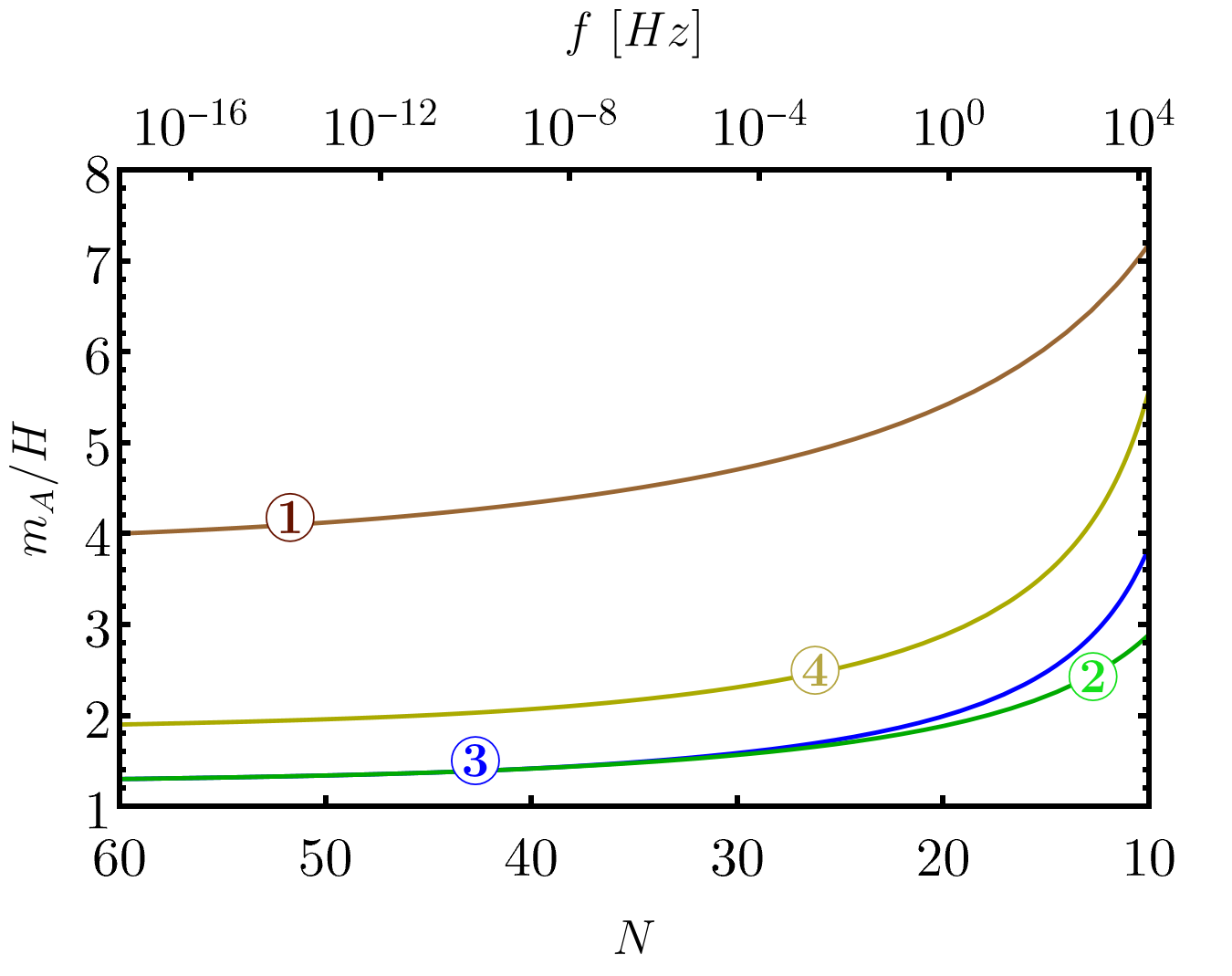} 
    }
    \caption{Evolution of model parameters $\xi$ and $m_A/H$ for four benchmark points: \Circled{1} $m_A=4H, \xi_C = 4.5$, \Circled{2} $m_A=1.3H, \xi_C=2.75$, \Circled{3} $m_A=1.3H, \xi_C=2.5$, \Circled{4} $m_A=1.9H, \xi_C=2.75$ in the context of the Starobinsky model. 
    See text for details.}
    \label{fig:evolwithN}
\end{figure}
Solving eqs.~\eqref{phiEOM} and \eqref{HEOM} numerically for a given potential, we get $H(N)$ and $\phi(N)$, which can be used to yield $\xi(N) \equiv d\phi/dN / (2\Lambda)$.

As a specific example we adopt the generalized Starobinsky model \cite{Starobinsky:1980te} which is a promising model with respect to the spectral index, $n_s$ vs. tensor-to-scalar ratio, $r$ plot from combined Planck 2018 analysis \cite{Planck:2018jri}. The inflaton potential in this model is given by
\begin{align}
    V(\phi) &= \frac{3}{4}V_0 \left[ 1-e^{-\gamma \phi} \right]^2,
\end{align}
where $V_0$ and $\gamma$ are free parameters, which can be constrained from CMB measurements of $n_s = 0.9649 \pm 0.0042$ (at $68\%$ CL) and $r < 0.056$ (at $95\% CL$) \cite{Planck:2018jri}. We choose $\gamma^2 = 8/125$ and $V_0 \approx 1.6 \times 10^{-9}$. In App. \ref{model_parameters} we justify the choice of these parameters. The evolution of $\xi$ and $m_A/H$ as a function of $N$ are shown in \cref{fig:evolwithN} for four benchmark points. These points are chosen because they will be used later to illustrate gravitational wave signals sensitive to various interferometers.

We choose $\xi$ and $m_A/H$ for all benchmark points at the CMB scale ($N\simeq60$) such that they are in the standard slow-roll regime where backreaction effects can be neglected. Initially $\xi$ increases rapidly until $N \sim 30-40$, when backreaction effects start to slow down its rise. Near the end of inflation backreaction becomes so severe that slow-roll condition is again established and $\xi$ rises swiftly. On the other hand, Hubble rate $H$ experiences a rather mild and monotonic decrease as $N$ decreases.

\section{Phenomenological Constraints at CMB Scale} \label{sec:5}

In this section we relate the $n$-point correlation functions computed in section~\ref{sec:4} to phenomenological observables at the CMB scale. In appropriate cases, we constrain the model parameter space from observational results.

\subsection{Scalar Power Spectrum}
In the absence of gauge field production, the scalar power spectrum is contributed by the usual vacuum fluctuations, and is given by
\begin{align}
    P_\zeta^{[\phi]} \equiv \left(\frac{H}{\dot{\phi}_0}\right)^2 \left(\frac{H}{2\pi} \right)^2. \label{scalarPinf}
\end{align}
Massive gauge field production facilitates inverse decay of the gauge bosons and gives rise to a second contribution proportional to the two-point correlation function computed in \cref{sec:3},
\begin{align}
   P_\zeta^{[A]} \equiv \frac{2 k^3}{(2\pi)^2}\langle\zeta_{\mathbf{k_1}}(\tau_0)\zeta_{\mathbf{k_2}}(\tau_0)\rangle_{(1)}^\prime.  
   \label{PzetaA}
\end{align}
The total scalar power spectrum combines these two effects
\begin{align}
    P_\zeta = P_\zeta^{[\phi]} + P_\zeta^{[A]}. \label{Pscalartotal}
\end{align}
Note that the two-point correlation function appearing in  \cref{PzetaA} is given by the in-in formalism  \cref{2pcfinin}, 
and it depends quadratically on $P_\zeta^{[\phi]}$, and on the model parameters $\xi$ and $m_A$.

\begin{figure}[!ht] 
    \centering
      \includegraphics[width=0.7\textwidth]{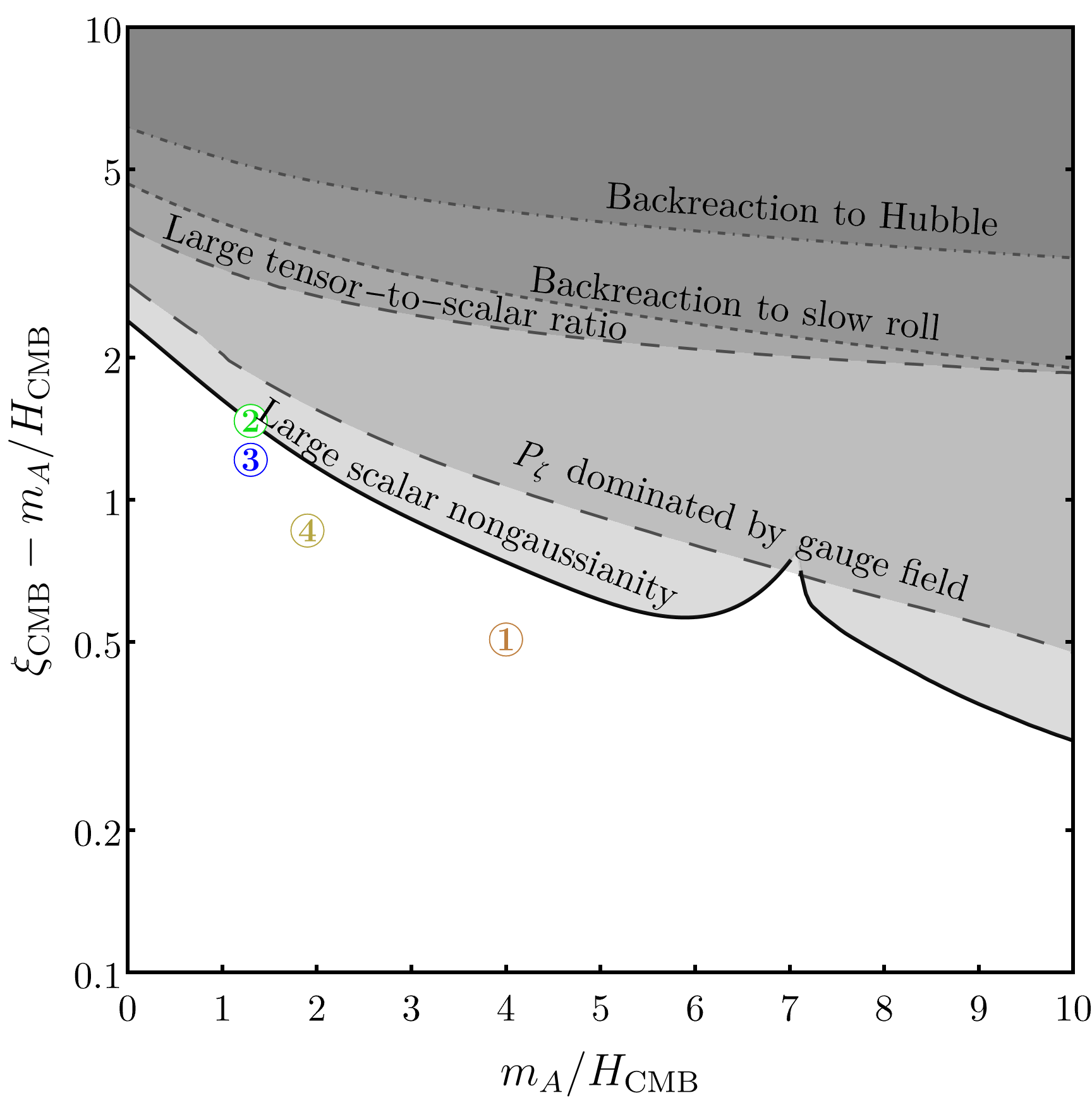}
    \caption{Shaded regions denote exclusion of the massive gauge boson's parameter space from various constraints. Circled numbers represent the benchmark points listed in Table \ref{table:BP}. Note that benchmark point \Circled{2} is very close to the upper bound set by scalar non-Gaussianity, but does not exceed it. See text for details. } \label{fig:AllConstraints}
\end{figure}

The amplitude of the scalar power spectrum at CMB scale is well measured \cite{WMAP:2010qai, Bunn:1996py},
\begin{align}
    P_\zeta \simeq 2.5 \times 10^{-9}, \label{PCOBE}
\end{align}
which accounts for the contribution of the inflaton as well as the extra degrees of freedom (massive gauge modes in this case), given by \cref{Pscalartotal}. If we make a conservative assumption that the gauge field's contribution is subdominant at the CMB, we can ignore $P_\zeta^{[A]}$ and fix $P_\zeta^{[\phi]} = 2.5 \times 10^{-9}$. This assumption would be valid as long as $P_\zeta^{[A]} \ll P_\zeta^{[\phi]}= 2.5 \times 10^{-9}$. In fig.~\ref{fig:AllConstraints}, we show the parameter space where this is violated by the label ``$P_\zeta$ dominated by gauge field''. Also, this assumption should be satisfied unless the curvature perturbation will have 
too large non-Gaussianity to be consistent with the CMB observations. 

\subsection{Scalar Non-Gaussianity} 
The curvature perturbations generated by the gauge field are non-Gaussian, and can be studied through the three-point correlation function. 
The three-point correlation function can take a general form
\begin{equation}
    \langle\zeta_{\mathbf{k_1}}(\tau)\zeta_{\mathbf{k_2}}(\tau)\zeta_{\mathbf{k_3}}(\tau)\rangle'
      =   
      \frac{3}{10}    (2\pi )^{4}    
     P_\zeta^2  \, \frac{\textstyle \sum_i  k_i^3 } { \textstyle \prod_i  k_i^3 }  
    {\cal S} ({\bf k}_1, {\bf k}_2,  {\bf k}_3) . 
   \label{eq:3pointShape}
\end{equation}
For the equilateral shape ($k_1 = k_2 = k_3 $) non-Gaussianity, we express it as a dimensionless parameter
\begin{align}
    f_{\rm NL}^{\rm eq} &= \frac{10}{9 } \frac{k_1^6 }{ (2\pi)^{4}} \frac{\langle\zeta_{\mathbf{k_1}} \zeta_{\mathbf{k_2}} \zeta_{\mathbf{k_3}}\rangle'}{P_\zeta(k)^2 }.
\end{align}
In fig.~\ref{fig:AllConstraints}, we show the parameter space violating the Planck 2018 constrain on equilateral non-Gaussianity $f_{\rm NL}^{\rm eq} = -25 \pm 47$ at $68\%$ CL \cite{Planck:2019kim}. This is more stringent than the other constraints we consider in this section. Note that left part of the bound is related to $f_{\rm NL} > -25+47$ whereas the right part corresponds to $f_{\rm NL} > -25-47$.




%
\subsection{Tensor Power Spectrum}
In the absence of massive gauge field production, the tensor power spectrum contributed by the usual vacuum fluctuations and is given by
\begin{align}
    P_h^{[\phi]} = \frac{2}{\pi^2}\left(\frac{H}{M_{\rm Pl}}\right)^2.
\end{align}
Similar to the scalar power spectrum, the contribution of the gauge field induced tensor perturbations to the power spectrum is given by
\begin{align}
    P_{h}^{[A],\pm} = \frac{2 k^3}{(2\pi)^2} \langle h^\pm(\tau_0, {\mathbf{k}_1}) h^\pm(\tau_0, -{\mathbf{k}_1}) \rangle_{(1)}. \label{tensorpowerA}
\end{align}
$\pm$ corresponds to the two polarizations of the graviton. The final expression for the two-point correlation function in \cref{tensorpowerA} in given in \cref{2PCFtensor}.

The total power spectrum is expressed as
\begin{align}
    P_h = \left[\frac{1}{\pi^2} \left( \frac{H}{M_{\rm Pl}} \right)^2 + P_h^{[A],+}\right] + \left[\frac{1}{\pi^2} \left( \frac{H}{M_{\rm Pl}} \right)^2 + P_h^{[A],-}\right] 
    &= P_h^+ + P_h^-, \label{PtensorTotal}
\end{align}
where we have included equal parts of the vacuum contribution to the two polarizations.

The power spectrum is chiral because of the parity-violating Chern-Simons interaction $\phi F \tilde{F}$. In the calculation of the two-point correlation function, this enters through the polarization vector contractions in \cref{inin2pcftensor}. An intuitive understanding develops by taking the $|\mathbf{k_1} - \mathbf{p_1}| \simeq |\mathbf{k_1}|$ limit in this equation. In this case the $\lambda = -$ mode vanishes whereas the $\lambda = +$ mode survives, since ${\boldsymbol{\epsilon}_{+}}(\mathbf{k_1}) \cdot {\boldsymbol{\epsilon}_{+}}(\mathbf{k_1}) = 0$ but ${\boldsymbol{\epsilon}_{+}}(\mathbf{k_1}) \cdot {\boldsymbol{\epsilon}_{-}}(\mathbf{k_1}) = 1$.

Even though gravitational waves generated from the tensor power spectrum has not been detected at the CMB scale, there are strict constraints on the ratio of tensor power spectrum to scalar power spectrum. This parameter, dubbed as tensor-to-scalar ratio, is defined as
\begin{align}
    r \equiv \frac{P_h}{P_\zeta} = \frac{P_h^+ + P_h^-}{2.5 \times 10^{-9}}. \label{r}
\end{align}
Combining the latest Planck 2018 data with the BICEP-Keck data constrains this at $r_* \leq 0.056$ at the CMB scale \cite{Planck:2018jri}. The region where this is violated is shown in fig.~\ref{fig:AllConstraints} with the label ``Large tensor-to-scalar ratio'', assuming $H/M_{\rm Pl} = 10^{-5}$.\footnote{Note that if the tensor power is dominated by the inflaton's contribution, $P_h^{[\phi]}$, $r<0.056$ implies $\frac{H}{M_{\rm Pl}} < 5.26 \times 10^{-5}$.} Noticeably, this is weaker than the constraints from scalar perturbations. For smaller $H/M_{\rm Pl}$ the bound from tensor-to-scalar ratio would be further weakened.


\subsection{Tensor Non-Gaussianity}
Similar to the scalar case, we can define equilateral $f_{\rm NL}$ for tensor perturbations. The current bound on tensor non-Gaussianities at CMB scales are much weaker than scalar non-Gaussianities. The most stringent bound comes from Planck T+E, $f_{\rm NL} < 800 \pm 1100$ \cite{Planck:2019kim}. It is much relaxed than the other bounds we have discussed in this section and is not shown in fig.~\ref{fig:AllConstraints}.

\subsection{Backreaction to Hubble Rate and Slow Roll}
The regions where the conditions \eqref{BR2} and \eqref{BR1} for negligible backreaction at CMB scales  are violated are labeled as ``Backreaction to Hubble'' and ``Backreaction to slow roll'', respectively, in fig.~\ref{fig:AllConstraints}. In deriving these boundaries, we have used $H/M_{\rm Pl} = 10^{-5}$ and $P_\zeta^{[\phi]} = 2.5 \times 10^{-9}$, which warrants some clarification. If $H/M_{\rm Pl}$ is smaller, the first bound would be weakened. On the other hand, the second bound is necessarily weaker as $P_{\zeta}^{[\phi]} \ll P_{\zeta}^{[A]}$ already at the boundary. 

The fact that the backreaction bounds are relaxed compared to the constraints from CMB observables ensures that in the allowed region backreaction effects are negligible. For the rest of the paper, we will only consider this allowed region at the CMB scales. For observables at smaller scales, we will calculate the evolution of the parameters $\xi$ and $m_A/H$ considering the backreaction effects.


\section{Gravitational Wave Signatures} \label{sec:7}

The tensor perturbations sourced by the massive gauge field left the horizon during inflation. Upon horizon re-entry, the power spectrum of the tensor perturbations can source gravitational waves whose amplitude today is given by
\begin{align}
    \Omega_{GW}(f) &\equiv \frac{1}{24}\Omega_{R,0} P_h(f). \label{OmegaGW}
\end{align}
Here $\Omega_{R,0} \simeq 8.6 \times 10^{-5}$ denotes the radiation energy density today and $P_h(f)$ is the frequency dependent power spectrum of the tensor fluctuations at the time of horizon exit. In our calculation of the tensor power spectrum we use the exact expression derived from in-in formalism for greater accuracy.

The power spectrum depends on the model parameters $\xi$ and $m_A/H$, whose evolution with efolding number $N$ was discussed in \cref{sec:3}. The frequency dependence can then be incorporated by employing the relation between $N$ and frequency $f$ \cite{Domcke:2016bkh}
\begin{align}
    N = N_{\rm CMB} + \log{\frac{k_{\rm CMB}}{0.002\ \rm{Mpc}^{-1}}} - 44.9 - \log{\frac{f}{10^2\ \rm{Hz}}}. \label{Nandf}
\end{align}
Typically $k_{\rm CMB} = 0.002\ \rm{Mpc}^{-1}$ and $N_{\rm CMB} \sim 50-60$.

For lower frequencies near the CMB scales, the effect of the gauge field creation on the tensor fluctuations is minimal even for large Hubble rate, and the power spectrum is dominated by the vacuum fluctuations. Current bound on scale-invariant stochastic gravitational wave at the CMB scales implies a tensor-to-scalar ratio $r < 0.056$ \cite{Planck:2018jri}, which gives $H/M_{\rm Pl} \lesssim 2.6 \times 10^{-5}$, and $\Omega_{\rm GW} < 1.2 \times 10^{-16}$. Such small gravitational wave amplitudes are only sensitive to planned interferometers DECIGO \cite{Kudoh:2005as, Kawamura:2020pcg} and BBO \cite{Harry:2006fi}, which are not expected to be operational in the next decade.  

Larger frequencies correspond to modes which left the horizon later than the CMB modes. By that time the rolling speed of the inflaton increases and the Hubble rate decreases, the combined effect of which implies a larger chemical potential. This dramatically enhances the power spectrum of the tensor perturbations sourced by the gauge field and it quickly supersedes the contribution from the vacuum fluctuations. Gravitational wave amplitude that eludes observation at the CMB scale now offers the possibility of detection at the interferometer scales.

There are roughly three frequency bands which are currently being probed (or are planned to be probed) by currently operational (future) interferometers. In the nanoHz range ($10^{-9} - 10^{-7}$ Hz), pulsar timing arrays (PTA) EPTA and NANOGrav are currently operating and have set upper bounds on the stochastic gravitational wave background.\footnote{NANOGrav has potentially detected a signal, but the source of the signal is still not unanimously known \cite{NANOGrav:2020bcs}.} In the same band, there are planned PTAs, SKA \cite{Janssen:2014dka} and IPTA \cite{Verbiest:2016vem}, with much higher sensitivity. The next band is mHz to Hz which will be investigated by planned laser interferometers LISA \cite{LISA:2017pwj}, BBO \cite{Harry:2006fi} and DECIGO \cite{Kudoh:2005as, Kawamura:2020pcg} and atomic interferometers AION \cite{Badurina:2019hst} and AEDGE \cite{AEDGE:2019nxb}. Currently operational advanced LIGO  and VIRGO \cite{LIGOScientific:2022sts} are sensitive to the $100$ Hz band and have set an upper limit \cite{KAGRA:2021kbb, Jiang:2022uxp}. Their planned upgrades will increase their sensitivity by at least an order \cite{KAGRA:2021kbb}. In the same band, planned Einstein Telescope (ET) \cite{Hild:2008ng} will be able to probe signals three orders of magnitude weaker. 

In order to see the explicit frequency dependence of $P_h$, we note that both the Hubble rate, $H$, and the chemical potential, $\xi$, varies with frequency (or equivalently, efolding number $N$) as we discussed in section \ref{sec:3} assuming a Starobinsky potential for the inflaton potential.

\begin{figure}[!ht] 
    \centering
      \includegraphics[width=0.99\textwidth]{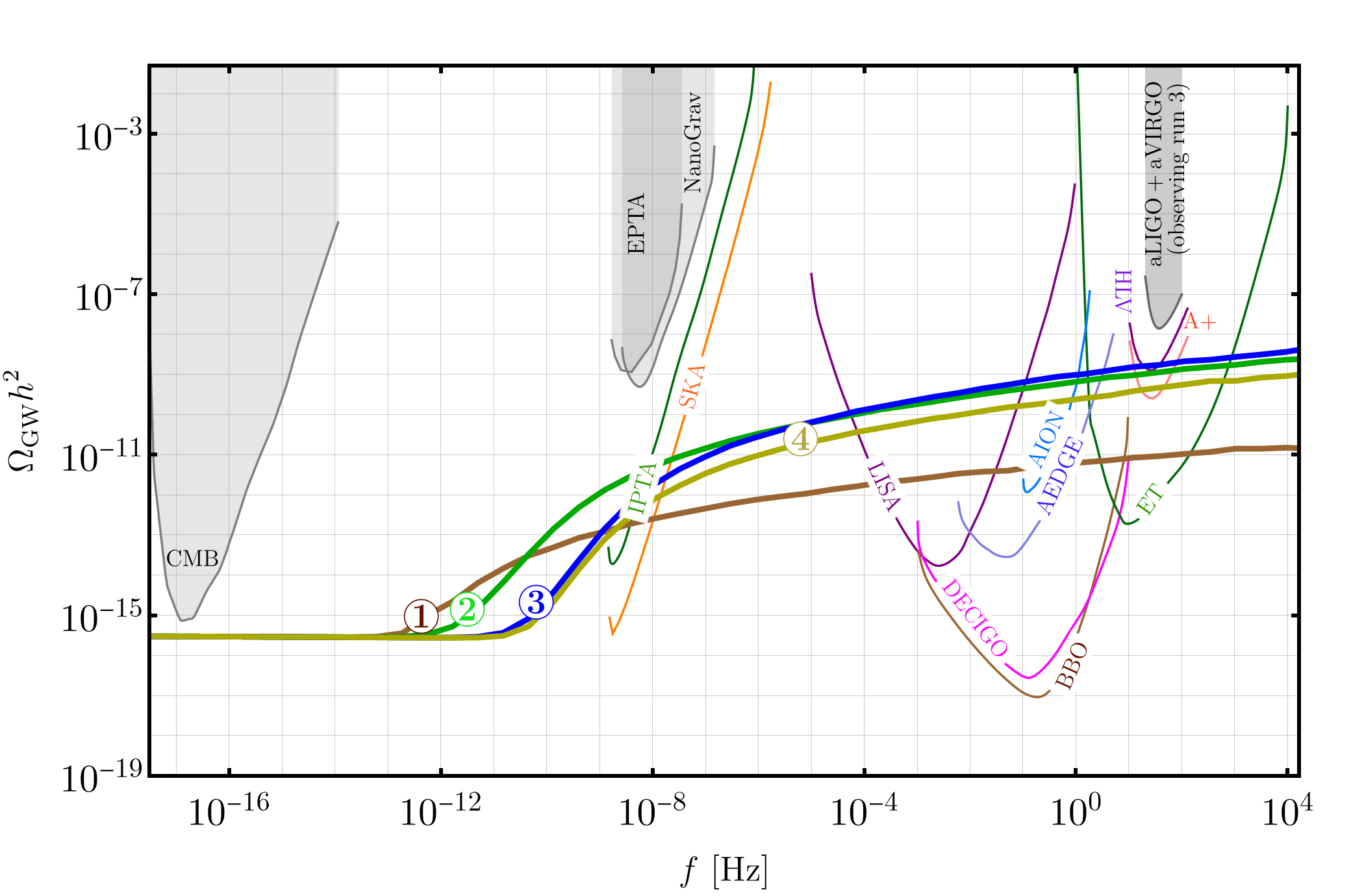}
    \label{fig:GWdetail}
    \caption{Gravitational wave spectrum for four benchmark points listed in Table \ref{table:BP} in the context of the generalized Starobinsky model. For comparison  we show the current upper bound (in gray) and future sensitivities (in color) of ongoing and proposed interferometers.
    See text for details.
    }
\end{figure}

Incorporating the variation of $\xi$ and $m_A/H$ with $f$, we show the gravitational wave amplitude $\Omega_{\rm GW}h^2$ ($h=0.7$) as a function of frequency for four benchmark points (listed in Table \ref{table:BP}) in fig.~\ref{fig:GWdetail}. In all cases, we notice that the gravitational wave amplitude is dominated by the vacuum fluctuations near CMB scales, before they rise at higher frequencies. 
\begin{table}[h!]
\centering
{\renewcommand{\arraystretch}{1.3}
\begin{tabular}{c@{\hskip 0.3in}c@{\hskip 0.5in}c}
    \toprule

  Benchmark Point  & ${m_A}/{H_{\rm CMB}}$ & $\xi_{\rm CMB}$ \\
 \midrule
 \Circled{1} & $4$ & $4.5$ \\
 \Circled{2} & $1.3$ & $2.75$  \\
\Circled{3} & $1.3$ & $2.5$  \\
\Circled{4} & $1.95$ & $2.75$  \\
\bottomrule
\end{tabular}
}
\caption{Benchmark points for gravitational wave signals.}
\label{table:BP}
\end{table}

Let us take benchmark point \Circled{3} as our main example, while the other points illustrate how the signal may vary with respect to the model parameters. The signal \Circled{3} rises early enough to be sensitive to IPTA, continues to be sensitive to a wide band of LISA, evades the upper bound set by LIGO+VIRGO but remains sensitive to their planned upgrades. This embodies the main characteristic of gravitational wave signals generated by massive gauge fields produced through the $\phi F \tilde{F}$ interaction --- low-lying signals undetectable at CMB scales rising at larger frequencies to be probed in a wide range of ground- and space-based interferometers.   

Benchmark points \Circled{2} and \Circled{4} demonstrate how this signal depends on the two parameters $\xi$ and $m_A/H$ at the CMB scales. Compared to \Circled{3}, $\xi$ is higher in \Circled{2} keeping $m_A/H$ unchanged. As expected, a higher chemical potential makes the contribution of the gauge field larger, and the signal surpasses the vacuum contribution earlier. However, at larger frequencies, backreaction effects also become stronger as seen in fig.~\eqref{fig:xivsN}, so much so that this signal goes slightly below \Circled{3}. On the other hand, at the CMB scale benchmark point \Circled{4} has the same $\xi$ as \Circled{2}, while its $m_A/H$ is larger. Heavier particles are less abundantly produced, and it takes longer for the gauge field contribution to dominate the vacuum contribution. In general, the signal for \Circled{4} remains slightly weaker for all observable frequencies. Finally, benchmark point \Circled{1} shows what happens when $\xi$ is larger compared to the previous three points. In this case, various constraints shown in fig.~\ref{fig:AllConstraints} dictate a reasonable choice of $m_A/H$. As expected, the signal starts to rise from the CMB level earlier than others, but severe backreaction effects weakens it in higher frequencies. This signal remains sensitive IPTA and LISA, but not to planned upgrades of LIGO+VIRGO. It, however, can be probed at ET in the same frequency band. We have checked that increasing $m_A/H$ and choosing a permissible $\xi$ from fig.~\ref{fig:AllConstraints} yields gravitational wave signals further suppressed compared to \Circled{1}.

A qualitative understanding for the suppression of the signals for larger $m_A/H$ can be obtained as follows. Massive particle production is restricted by the Boltzmann suppression factor $e^{-\pi m/H}$, while in the case of gauge fields produced from  $\phi F \tilde{F}$ interaction, is enhanced by the factor $e^{\pi \xi}$. Therefore, the overall strength of the signals roughly depends on $\xi - m_A/H$. From the constraint plot fig.~\ref{fig:AllConstraints}, larger $m_A/H$ has a smaller upper bound for allowed $\xi-m_A/H$. This implies that, within the allowed parameter space, larger $m_A/H$ would eventually yield a weaker signal at interferometer scales, especially at LISA and LIGO+VIRGO scales. 

We now briefly comment about the effect of reheating on the gravitational wave spectrum. The amplitude of the gravitational wave depends on the details of the reheating history and is typically suppressed for a matter dominated reheating phase \cite{Turner:1993vb, Seto:2003kc, Nakayama:2008ip, Buchmuller:2013lra}.\footnote{Assuming radiation domination after reheating, gravitational wave production can be too strong in the case of massless gauge bosons \cite{Adshead:2018doq, Adshead:2019lbr, Adshead:2019igv}. The case of massive gauge bosons is yet to be explored.} Eq.~\eqref{OmegaGW} has been derived assuming instantaneous reheating (equation of state $\omega = 1/3$) and further assuming that the degrees of freedom of the thermal bath remained unchanged between horizon re-entry and today. The energy density of the gauge field becomes comparable to the vacuum density near the end of inflation, suggesting a matter dominated era, so that the equation of state should have an intermediate value between that of radiation and matter \cite{Podolsky:2005bw}. Furthermore, it modifies \cref{Nandf} introducing a term dependent on reheating temperature and may imply a shorter duration of inflation \cite{Liddle:1993fq, Liddle:2000cg}. The combined effect is a possible suppression of the spectrum for frequencies larger than $f_{\rm rh} \simeq 0.4$ Hz ($T_{\rm rh}/10^7$ GeV), where $T_{\rm rh}$ denotes the reheating temperature \cite{Turner:1993vb, Seto:2003kc, Nakayama:2008ip}, which may hide a potential signal from the LIGO band, but typically not from other interferometers located at lower frequencies.

\begin{figure}[!ht]
    \centering
    \includegraphics[width=0.65\textwidth]{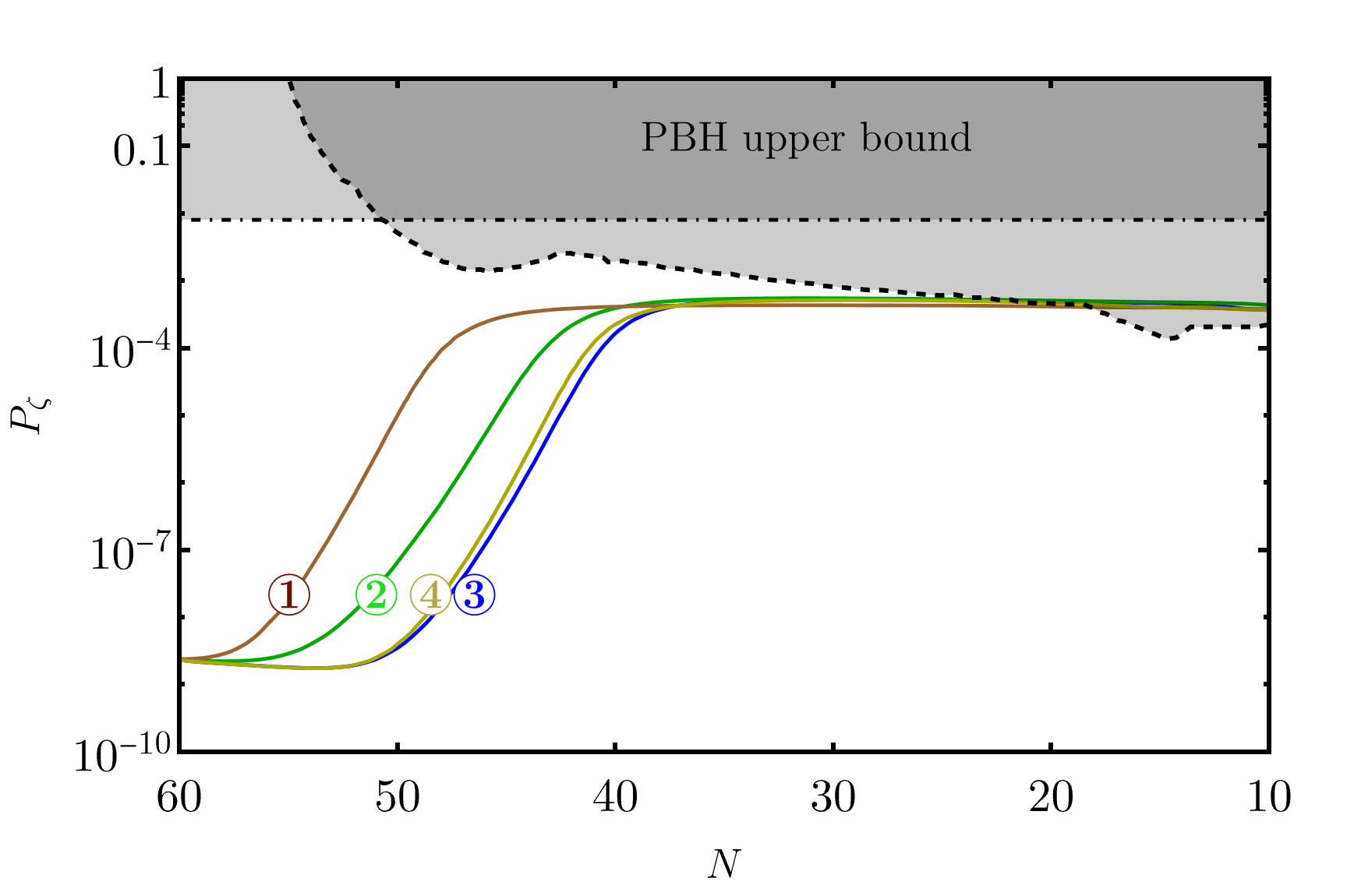}
    \caption{Evolution of the scalar power spectrum in the context of the generalized Starobinsky model for the same benchmark points as in fig.~\ref{fig:GWdetail}. Gray area represents overproduction of primordial black holes. The upper line corresponds to gaussian perturbations and the lower curve corresponds to non-Gaussian perturbations. See text for details.}
    \label{fig:PScalarBR}
\end{figure}

Finally, we discuss about the possibility of primordial black hole (PBH) creation from excessive scalar perturbation in the context of the generalized Starobinsky model. The non-observation of PBH sets strong constraints on the fraction of energy going into PBHs at their creation as a function of PBH mass. PBH masses below $10^{15}$ g can be detected from their entropy production in the early universe, masses around $10^{15}$ g can be detected from signals in $\gamma$-rays, and heavier masses stable PBHs can be searched for in lensing experiments \cite{Carr:2009jm}. The mass of a PBH can be associated with the efolding number $N$ when the perturbation sourcing the creation of the PBH left the horizon. Following the estimates in Refs.~\cite{Josan:2009qn, Carr:2020gox}, an upper bound on the scalar power spectrum as a function of $N$ was presented in Ref.~\cite{Linde:2012bt}. This is shown in \cref{fig:PScalarBR} with a dashed curve, where we also show the the evolution of the scalar power spectrum considering strong backreaction in the context of the generalized Starobinsky model.\footnote{Strong backreaction introduces an extra term in the equation of motion of the inflaton perturbation, see \cref{EOMscalarperturb2}. The scalar spectrum curves shown in \cref{fig:PScalarBR} have been derived using an approximate formula following the technique of Ref.~\cite{Linde:2012bt}.}

Note that the bound derived in Ref.~\cite{Linde:2012bt} has an $\mc O(1)$ uncertainty because of the approximations involved in the calculation. Our benchmark points violate this bound only at high frequencies by $\mc O(1)$. Furthermore, in recent literature this bound has been debated from various considerations. In deriving this bound, Ref.~\cite{Linde:2012bt} assumed that the curvature perturbation is non-Gaussian and can be expressed as
\begin{align}
    \zeta = g^2 - \langle g^2\rangle,
\end{align}
where $g$ follows a gaussian distribution. Consequently, the probability distribution function of $\zeta$ can be derived from $P(\zeta)d\zeta = P(g) dg$, and follows a chi-squared distribution
\begin{align}
    P(\zeta) = \frac{1}{\sqrt{2\pi(\zeta + \sigma^2)\sigma}} e^{-\frac{\zeta+\sigma^2}{2\sigma^2}},
\end{align}
with $\sigma^2 \equiv \langle g^2\rangle$. A recent lattice study \cite{Caravano:2022epk} shows that at smaller scales the curvature perturbation actually becomes nearly gaussian because of the strong backreaction from gauge field production. A plausible explanation is, in the strong backreaction regime, large number of excited gauge modes are produced contributing to the source term $\mathbf{E}\cdot \mathbf{B}$, and central limit theorem dictates that their overall effect is gaussian. If the curvature perturbation follows a nearly gaussian statistics, the upper bound on primordial black hole overproduction is relaxed \cite{Linde:2012bt, Lyth:2012yp} 
\begin{align}
    P(\zeta) \lesssim 0.008 - 0.05.
\end{align}
We have shown the $P_{\zeta} > 0.008$ region in fig.~\ref{fig:PScalarBR} bounded with a dot-dashed line.
While our benchmark points violate the bound for non-Gaussian perturbations at high frequencies by $\mc O(1)$, assuming a reversion to gaussianity at those scales would relax the bound and potentially allow this model to avoid the overproduction of primordial black holes. 

Even if the bound derived assuming non-Gaussian perturbations remains valid at all scales, it can be avoided by introducing $\mc N$ copies of the gauge field.  In this case, the scalar power spectrum is reduced by a factor of $\mc N$ at small scales \cite{Anber:2009ua}. For the benchmark points shown in \cref{fig:PScalarBR}, $\mc N \sim 2-3$ would be sufficient to evade the PBH bound for non-Gaussian perturbations at high frequencies. This would also weaken the gravitational wave signals similarly, and might affect observability at LIGO scales.

We also note that the large scalar perturbations on small scales can also lead to sizable second order tensor perturbations \cite{Baumann:2007zm}, however, these are subdominant compared to the leading order gravitational wave contribution calculated above.


\section{Conclusion and Outlook} \label{sec:8}
In this paper we have shown that gravitational wave signals could be a complementary window into ``cosmological collider'' physics in the context of massive $U(1)$ gauge bosons. The presence of Hubble-scale massive particles generate an oscillatory signal at the scales of cosmic microwave background (CMB) or large scale structure (LSS) in the ``squeezed limit'' of the scalar bispectrum, where one of the external momenta in the three-point correlation function is much smaller compared to the other two. The frequency of this signal is proportional to the mass of the particle, thus divulging its presence in the ``cosmological collider''. We have extended the scope of discovering massive gauge bosons present during inflation to much smaller scales through the detection of characteristic gravitational waves generated by primordial tensor fluctuations sourced by these particles.

Massive gauge bosons can be efficiently produced during inflation from the decay of the inflaton due to a Chern-Simons coupling $\phi F \tilde{F}$.  Inverse decay of the gauge modes leave observable imprints on the primordial scalar and tensor fluctuations. We have derived the updated constraints on the parameter space of gauge boson production from various bounds at the CMB scales using latest cosmological data. We find that the scalar non-Gaussianity bound from Planck 2018 data puts the most stringent bound on the parameter space, essentially eliminating the dimensionless chemical potential larger than the mass to Hubble ratio by $0.5 - 1.5$. 

We then extrapolated the scalar and tensor power spectrum beyond CMB scales, consistently taking into account the backreaction effect of the massive gauge modes on the perturbations leaving the horizon at later stages of inflation. In the allowed parameter space, we have demonstrated in \cref{fig:GWdetail} that characteristic gravitational wave signals from tensor perturbations discoverable at current and planned interferometers emerge. These signals remain flat near CMB scales evading the stringent upper bound, but rises at smaller scales to become sensitive to the gravitational wave detectors. Furthermore, such signals span the entirety of the frequency bands probed by current and planned terrestrial and space-based gravitational wave detectors. Non-observation at multiple bands can easily rule out the model. 


\acknowledgments
We are grateful to Rodolfo Capdevilla, Yanou Cui, Rouven Essig, Jiamin Hou, Gordan Krnjaic, John March-Russell, Stefano Profumo, Subir Sarkar, Pierre Sikivie,  Zachary Slepian, Zhong-Zhi Xianyu, Haibo Yu, and Yi-Ming Zhong
for useful discussion.
This work was supported in part by the U.S. Department of Energy under grant DE-SC0022148 at the University of Florida. MHR acknowledges financial support from the STFC Consolidated Grant ST/T000775/1, and from Maurice
C. Holmes and Frances A. Holmes Endowed Fellowship. The authors acknowledge University of Florida Research Computing for providing computational resources and support that have contributed to the research results reported in this publication. 
\appendix

\section{Correlation Function Calculation in Dominant Real Mode Function Approximation} \label{App:A}
In this appendix we derive explicit expressions for the correlation functions introduced in section~\ref{sec:4} using the dominant real mode function approximation.
\subsection{Curvature Perturbation}
\subsubsection{Two-point correlation function}

Using \cref{2pcfinin2}, and simplifying the polarization vector contractions using
\begin{align}
    \left| \boldsymbol{\epsilon}_{+}(\mathbf{k_1}) \cdot \boldsymbol{\epsilon}_{+}(\mathbf{k_2}) \right|^2 = \frac{1}{4}\left(1 - \frac{\mathbf{k_1}\cdot \mathbf{k_2}}{k_1 k_2}\right)^2,
\end{align}
the two-point correlation function becomes
\begin{align}
     \langle\zeta_{\mathbf{k_1}}(\tau_0)\zeta_{\mathbf{k_2}}(\tau_0)\rangle' &= \frac{H^6}{8\Lambda^2 \dot{\phi}_0^2} \frac{1}{k^6}  \int \frac{d^3q_1}{(2\pi)^3} \left[1 - \frac{(\mathbf{k_1}-\mathbf{q_1})\cdot \mathbf{q_1}}{|\mathbf{k_1}-\mathbf{q_1}| q_1}\right]^2 \nonumber \\
     &\times \left| \int d\tau_1 (k \tau_1 \cos{k\tau_1} - \sin{k\tau_1}) \mc B_2(|\mathbf{k_1}-\mathbf{q_1}|,q_1;\tau_1) \right|^2,
\end{align}
where the $\delta$-function $(2\pi)^3 \delta^{(3)} (\mathbf{k_1}+\mathbf{k_2})$ has been stripped off.
For further simplification, we define the momentum $3$-vectors, without loss of generality, as follows:
\begin{align}
    \mathbf{k_1} &\equiv (0,0,k), \\
    \mathbf{q_1} &\equiv k q (0, -\sin{\theta}, \cos{\theta}).
\end{align}
After some straightforward algebra, the two-point correlation function can be expressed as
\begin{align}
    \langle\zeta_{\mathbf{k_1}}(\tau_0)\zeta_{\mathbf{k_2}}(\tau_0)\rangle' &= \frac{\pi^2}{2} { P_\zeta^{[\phi]}}^2 \xi^2 e^{2\pi\xi} \frac{1}{k^3} \int_{q=0}^{\infty} dq q^2  I_{\theta}^\zeta(q) \left| I_x^\zeta (\theta,q) \right|^2, 
\end{align}
where the usual scalar power spectrum contributed by the inflaton is defined as
\begin{align}
    P_\zeta^{[\phi]} \equiv \left(\frac{H}{\dot{\phi}_0}\right)^2 \left(\frac{H}{2\pi} \right)^2, 
\end{align}
and the integrals with respect to $\theta$ and $x \equiv -k\tau$ are defined as
\begin{align}
    I_{\theta}^\zeta(\theta, q) &\equiv \int_{\theta = 0}^{\pi} d\theta \sin{\theta} \left[ 1+\frac{q-\cos{\theta}}{\sqrt{1+q^2-2q\cos{\theta}}} \right]^2, \\
    I_x^\zeta(x, \theta, q) &\equiv \int_{x=0}^{\infty} dx (x\cos{x} - \sin{x}) \nonumber \\
    &\times \left[ \frac{\sqrt{q}}{(1+q^2-2q\cos{\theta})^{1/4}} \partial_x W_{-i\xi, i\mu}(-2ix\sqrt{1+q^2-2q\cos{\theta}}) W_{-i\xi, i\mu}(-2ixq) \right. \nonumber \\
    &\left. + \frac{(1+q^2-2q\cos{\theta})^{1/4}}{\sqrt{q}} \partial_x W_{-i\xi, i\mu}(-2ixq) W_{-i\xi, i\mu}(-2ix\sqrt{1+q^2-2q\cos{\theta}}) \right].
\end{align}

\subsubsection{Three-point correlation function}
In order to evaluate the polarization vector contractions, we first need to choose a suitable set of momentum vectors $\mathbf{k_1}$, $\mathbf{k_2}$, $\mathbf{k_3}$. Then, for any $3$-vector $\mathbf{v} \equiv v(\sin{\theta_\mathbf{v}} \cos{\phi_\mathbf{v}}, \sin{\theta_\mathbf{v}}\sin{\phi_\mathbf{v}}, \cos{\theta_\mathbf{v}})$, the polarization vector is defined as
\begin{align}
    \boldsymbol{\epsilon_\pm}(\mathbf{v}) \equiv \frac{1}{\sqrt{2}} \left(\mp \cos{\theta_\mathbf{v}} \cos{\phi_\mathbf{v}} + i\sin{\phi_\mathbf{v}}, \mp \cos{\theta_\mathbf{v}} \sin{\phi_\mathbf{v}} - i\cos{\phi_\mathbf{v}}, \pm \sin{\theta_\mathbf{v}} \right), \label{polarizationformula}
\end{align}
so that the properties in \cref{polarization} are satisfied.

Without loss of generality, we define the momentum $3$-vectors as
\begin{align}
    \mathbf{k_1} &\equiv k(0,0,1)\equiv k \hat{k}_1, \\
    \mathbf{k_2} &\equiv y_2 k (\sin{\theta_2}, 0, \cos{\theta_2}) \equiv y_2 k \hat{k}_2, \\
    \mathbf{k_3} &\equiv y_3 k (-\sin{\theta_3}, 0, \cos{\theta_3}) \equiv y_3 k \hat{k}_3.
\end{align}
Since $\mathbf{k_1} + \mathbf{k_2} + \mathbf{k_3} = 0$, the momentum vectors can be expressed in terms of three free parameters. Choosing these to be $k, y_2$ and $y_3$, we express $\theta_2$ and $\theta_3$ as
\begin{align}
    \theta_2 &= \pi - \cos^{-1}\left(\frac{1+y_2^2-y_3^2}{2y_2}\right), \\
    \text{and}\quad \theta_3 &= \pi - \cos^{-1}\left(\frac{1+y_3^2-y_2^2}{2y_3}\right).
\end{align}
Defining the loop momentum as 
\begin{align}
    \mathbf{q} &\equiv k \mathbf{p} \equiv kp(\sin{\theta} \cos{\phi}, \sin{\theta}\sin{\phi}, \cos{\theta}),
\end{align}
and using 
\begin{align}
    \theta_{\mathbf{v}} &\equiv \cos^{-1}{\left( \frac{v_z}{\sqrt{v_x^2+v_y^2+v_z^2}} \right)}, \\
    \text{and}\quad \phi_{\mathbf{v}} &\equiv \tan^{-1}{\left( \frac{v_y}{v_x} \right)} + \pi \Theta(-v_x)
\end{align}
for any $3$ -vector $\mathbf{v} \equiv (v_x, v_y, v_z) \equiv v(\sin{\theta_\mathbf{v}} \cos{\phi_\mathbf{v}}, \sin{\theta_\mathbf{v}}\sin{\phi_\mathbf{v}}, \cos{\theta_\mathbf{v}})$, where $\Theta(x)$ is the Heaviside theta function, we get
\begin{align}
    \theta_\mathbf{p} &= \theta, \label{theta_q}\\
    \phi_{\mathbf{p}} &= \tan^{-1}{[\tan{\phi}]} + \pi \Theta(-p \sin{\theta} \cos{\phi}),\\
    \theta_{\hat{k}_1-\mathbf{p}} &= \cos^{-1}{\left[ \frac{1-p\cos{\theta}}{\sqrt{1+p^2-2p\cos{\theta}}} \right]}, \\
    \phi_{\hat{k}_1-\mathbf{p}} &= \tan^{-1}{\left[\tan{\phi}\right]} + \pi \Theta(p \sin{\theta} \cos{\phi}), \\
    \theta_{y_3\hat{k}_3+\mathbf{p}} &= \cos^{-1}{\left[ \frac{y_3 \cos{\theta_3} + p\cos{\theta}}{\sqrt{y_3^2 + p^2 -2y_3 p(\sin{\theta} \cos{\phi} \sin{\theta_3}-\cos{\theta} \cos{\theta_3})}} \right]},\\
    \phi_{y_3\hat{k}_3+\mathbf{p}} &= \tan^{-1}{\left[ \frac{p\sin{\theta} \sin{\phi}}{-y_3 \sin{\theta_3}+ p \sin{\theta} \cos{\phi}} \right]} + \pi \Theta(y_3 \sin{\theta_3}-p \sin{\theta}\cos{\phi}). \label{phi_k3plusq}
\end{align}
Defining  $x_i \equiv -k_i \tau_i$,
the mode function combinations can be expressed as
\begin{align}
    \mc B_2(|\mathbf{k_1} - \mathbf{q}|, q; \tau_1) &=  -\frac{k}{2}e^{\pi\xi} \mc W_1(p, |\hat{k}_1-\mathbf{p}|;x_1),\\
    \mc B_2(|\mathbf{k_1}-\mathbf{q}|,\overline{|\mathbf{k_3}+\mathbf{q}|};\tau_2) &=  -\frac{y_2 k}{2}e^{\pi\xi} \mc W_2(|\hat{k}_1-\mathbf{p}|,|y_3\hat{k}_3+\mathbf{p}|;x_2),\\
    \mc B_2({q},|\mathbf{k_3} + \mathbf{q}|; \tau_3) &=  -\frac{y_3 k}{2}e^{\pi\xi} \mc W_3(p,|y_3\hat{k}_3+\mathbf{p}|;x_3),
\end{align}
where $\mc W_i(a,b;x_i)$ denotes the dependence on Whittaker functions and dimensionless momenta, and  are defined as
\begin{align}
    \mc W_1(p, |\hat{k}_1-\mathbf{p}|;x_1) &\equiv \sqrt{\frac{p}{|\hat{k}_1-\mathbf{p}|}} \partial_{x_1} W(-2ix_1 |\hat{k}_1-\mathbf{p}|) W(-2ix_1 p)  \nonumber \\
    &+ \sqrt{\frac{|\hat{k}_1-\mathbf{p}|}{p}} \partial_{x_1} W(-2ix_1 p) W(-2ix_1 |\hat{k}_1-\mathbf{p}|), \label{W1}\\
    \mc W_2(|\hat{k}_1-\mathbf{p}|,|y_3\hat{k}_3+\mathbf{p}|;x_2) &\equiv  \sqrt{\frac{|\hat{k}_1-\mathbf{p}|}{|y_3\hat{k}_3+\mathbf{p}|}} \partial_{x_2} W^*\left(-2i\frac{x_2}{y_2} |y_3\hat{k}_3+\mathbf{p}|\right) W\left(-2i\frac{x_2}{y_2} |\hat{k}_1 - \mathbf{p}|\right)   \nonumber \\
    &+ \sqrt{\frac{|y_3\hat{k}_3+\mathbf{p}|}{|\hat{k}_1-\mathbf{p}|}} \partial_{x_2} W\left(-2i\frac{x_2}{y_2} |\hat{k}_1 - \mathbf{p}|\right) W^*\left(-2i\frac{x_2}{y_2} |y_3\hat{k}_3+\mathbf{p}|\right), \label{W2}\\
    \mc W_3(p,|y_3\hat{k}_3+\mathbf{p}|;x_3) &\equiv  \sqrt{\frac{p}{|y_3\hat{k}_3+\mathbf{p}|}} \partial_{x_3} W\left(-2i\frac{x_3}{y_3} |y_3\hat{k}_3+\mathbf{p}|\right) W\left(-2i\frac{x_3}{y_3} p\right)  \nonumber \\
    &+ \sqrt{\frac{|y_3\hat{k}_3+\mathbf{p}|}{p}} \partial_{x_3} W\left(-2i\frac{x_3}{y_3} p\right) W\left(-2i\frac{x_3}{y_3} |y_3\hat{k}_3+\mathbf{p}|\right). \label{W3}
\end{align}
Combining everything, the three-point correlation function can be expressed as
\begin{align}
    \langle\zeta_{\mathbf{k_1}}(\tau)\zeta_{\mathbf{k_2}}(\tau)\zeta_{\mathbf{k_3}}(\tau)\rangle' &= \frac{ {P_\zeta^{[\phi]}}^3 \xi^3 e^{3\pi\xi} }{(2\pi)^{3}y_2^3 y_3^3} \frac{\delta^{(3)}(\mathbf{k_1}+\mathbf{k_2}+\mathbf{k_3})}{k^6}  \int {d^3 p} [\boldsymbol{\epsilon}_{+}(\mathbf{p})\cdot \boldsymbol{\epsilon}_{+}(\hat{k}_1-\mathbf{p})]\nonumber \\ &\times [\boldsymbol{\epsilon}_{+}(-y_3\hat{k}_3-\mathbf{p})\cdot \boldsymbol{\epsilon}_{+}(-\hat{k}_1+\mathbf{p})]\ [\boldsymbol{\epsilon}_{+}(-\mathbf{p})\cdot \boldsymbol{\epsilon}_{+}(y_3\hat{k}_3+\mathbf{p})] \nonumber \\
    &\times \int dx_1 \left(x_1 \cos{x_1} - \sin{x_1}\right) \mc W_1(p, |\hat{k}_1-\mathbf{p}|;x_1) \nonumber \\
    &\times \int dx_2 \left(x_2 \cos{x_2} - \sin{x_2}\right) \mc W_2(|\hat{k}_1-\mathbf{p}|,|y_3\hat{k}_3+\mathbf{p}|;x_2) \nonumber \\
    &\times \int dx_3 \left(x_3 \cos{x_3} - \sin{x_3}\right) \mc W_3(p,|y_3\hat{k}_3+\mathbf{p}|;x_3).
    \label{eq:3PCF_scalar_generic}
\end{align}

\subsubsection{Squeezed limit}
The ``squeezed limit'' is defined as $|\mathbf{k_1}| \approx |\mathbf{k_2}| \gg |\mathbf{k_3}|$. Since $\mathbf{k_1}+\mathbf{k_2}+\mathbf{k_3} = 0$, without loss of generality, we choose
\begin{align}
    &\mathbf{k_1} \approx -\mathbf{k_2} \equiv k (0,0,1) \equiv k \hat{k}_1, \label{k12sq}\\
    &\mathbf{k_3} \equiv y_3 k (1,0,0) \equiv y_3 k \hat{k}_3 \label{k3sq}
\end{align}
with $y_3 \ll 1$. In this limit,  the dominant contribution of the three-point correlation function comes from the case $|y_3 \hat{k}_3 + \mathbf{p}| = |y_3 \hat{k}_3| = |\mathbf{p}|$, 
and the integral over loop momentum can be approximated as \cite{Chen:2018xck}
\begin{align}
    \int d^3 p &\approx y_3^3 \int d\phi. \label{sqcond}
\end{align}
Since $|\hat{k}_1 - \mathbf{p}| \approx 1$ and $|y_3 \hat{k}_3 + \mathbf{p}| \approx y_3$, the arguments of $\mc W_i$ functions are not dependent on $\phi$, and are given by
\begin{align}
    \mc W_1 (y_3,1;x_1) &= \sqrt{y_3} \partial_{x_1} W(-2ix_1) W(-2ix_1 y_3) + \frac{1}{\sqrt{y_3}} \partial_{x_1} W(-2ix_1y_3) W(-2ix_1),\\
    \mc W_2(1,y_3;x_2) &= \sqrt{y_3} \partial_{x_2} W(-2ix_2) W^*(-2ix_2 y_3) + \frac{1}{\sqrt{y_3}} \partial_{x_2} W^*(-2ix_2y_3) W(-2ix_2), \\
    \mc W_3(y_3,y_3;x_3) &= \partial_{x_3} W^2(-2ix_3).
\end{align}
Hence the angular integral can be performed over the polarization part only.
Furthermore, for $|y_3 \hat{k}_3 + \mathbf{p}| = |y_3 \hat{k}_3| = |\mathbf{p}|$, the loop momentum can be defined as
\begin{align}
    \mathbf{p} = y_3\left(-\frac{1}{2},\frac{\sqrt{3}}{2}\cos{\phi},\frac{\sqrt{3}}{2}\sin{\phi}\right). \label{psq}
\end{align}
Under this approximation, we have $\hat{k}_1 - \mathbf{p} \approx \hat{k}_1$, and the polarization vector dot products become
\begin{align}[\boldsymbol{\epsilon}_{+}(\mathbf{p})\cdot \boldsymbol{\epsilon}_{+}(\hat{k}_1)] [\boldsymbol{\epsilon}_{+}(-y_3\hat{k}_3-\mathbf{p})\cdot \boldsymbol{\epsilon}_{+}(-\hat{k}_1)] [\boldsymbol{\epsilon}_{+}(-\mathbf{p})\cdot \boldsymbol{\epsilon}_{+}(y_3\hat{k}_3+\mathbf{p})]. \label{poldot}
\end{align}
Following \cref{polarizationformula}, these polarization vectors are given by
\begin{align}
    \boldsymbol{\epsilon}(\hat{k}_1) &= \frac{1}{\sqrt{2}}(1,i,0),\\
    \boldsymbol{\epsilon}(\mathbf{p}) &= \frac{1}{\sqrt{2}}\left(-\frac{\sqrt{3}}{2},-\frac{1}{2}\cos{\phi}-i\sin{\phi},-\frac{1}{2}\sin{\phi}+i\cos{\phi}\right),\\
    \boldsymbol{\epsilon}(y_3\hat{k}_3+\mathbf{p}) &= \frac{1}{\sqrt{2}}\left(-\frac{\sqrt{3}}{2},\frac{1}{2}\cos{\phi}-i\sin{\phi},\frac{1}{2}\sin{\phi}+i\cos{\phi}\right).
\end{align}
Using these, the polarization vector dot products in \eqref{poldot} yield
\begin{align}
     &\frac{3\pi}{32}\left(\sqrt{3}+i\cos{\phi}-2\sin{\phi}\right)^2. 
\end{align}
Since $|\hat{k}_1 - \mathbf{p}| \approx 1$ and $|y_3 \hat{k}_3 + \mathbf{p}| \approx y_3$, the arguments of $\mc W_i$ functions are not dependent on $\phi$. Performing the angular integral  over the polarization part only, we get
\begin{align}
    \int_0^{2\pi} d\phi\ \frac{3\pi}{32}\left(\sqrt{3}+i\cos{\phi}-2\sin{\phi}\right)^2 = \frac{27 \pi^2}{32},
\end{align}
and the three point correlation function in the squeezed limit becomes
\begin{align}
    \langle\zeta_{\mathbf{k_1}}(\tau_0)\zeta_{\mathbf{k_2}}(\tau_0)\zeta_{\mathbf{k_3}}(\tau_0)\rangle' &= \frac{27}{256 \pi} {P_\zeta^{[\phi]}}^3 \xi^3 e^{3\pi\xi} \frac{\delta^{(3)}(\mathbf{k_1}+\mathbf{k_2}+\mathbf{k_3})}{k^6} \nonumber \\
    &\times \int dx_1 \left(x_1 \cos{x_1} - \sin{x_1}\right) \mc W_1(y_3, 1;x_1) \nonumber \\
    &\times \int dx_2 \left(x_2 \cos{x_2} - \sin{x_2}\right) \mc W_2(1,y_3;x_2) \nonumber \\
    &\times \int dx_3 \left(x_3 \cos{x_3} - \sin{x_3}\right) \mc W_3(y_3,y_3;x_3). \label{3PCFscalarsq}
\end{align}

\subsection{Tensor Perturbation}
\subsubsection{Two-point correlation function}
In the dominant real mode function approximation, \cref{inin2pcftensor} becomes
\begin{align}
    \left\langle h^{\lambda}(\tau_0, \mathbf{{k}_1})h^{\lambda}(\tau_0, \mathbf{{k}_2})\right\rangle' &= \frac{H^2}{M_{\rm Pl}^2}  \frac{2^2}{k^3} \int\frac{d^3\mathbf{p}}{(2\pi)^3}   \left|\boldsymbol{\epsilon}_{-\lambda}(\mathbf{k}_1)\cdot\boldsymbol{\epsilon}_+(\mathbf{p})\right|^2\ \left|\boldsymbol{\epsilon}_{-\lambda}(\mathbf{k}_1)\cdot\boldsymbol{\epsilon}_+(\mathbf{k}_1-\mathbf{p})\right|^2\nonumber\\ &\times \int_{-\infty}^0 d\tau_2 \int_{-\infty}^{0} d\tau_1\ A'(\tau_1, p)A'(\tau_1, {|\mathbf{k}_1-\mathbf{p}|})  A'^{*}  (\tau_2, p)A'^{*}(\tau_2, {|\mathbf{k}_1-\mathbf{p}|}) \nonumber\\ &\times  {\rm{Im}}\ h^{\lambda}_{{k}_1}(\tau_1)\  {\rm{Im}}\ h^{\lambda}_{{k}_1}(\tau_2). \label{inin2pcftensor3}
\end{align}
The polarization vectors can be simplified using
\begin{align}
    \left| \boldsymbol{\epsilon}^{(+)} (\mathbf{q_1}) \boldsymbol{\epsilon}^{(-\lambda)} (\mathbf{q_2}) \right|^2 = \frac{1}{4}\left(1+\lambda \frac{\mathbf{q_1}\cdot \mathbf{q_2}}{q_1 q_2}\right)^2.
\end{align}
Defining the momentum 3-vectors as $\mathbf{k_1} \equiv (0,0,k)$ and $\mathbf{p_1} \equiv kq(0, -\sin\theta, \cos \theta)$, and changing variables to $x_i \equiv -k\tau_i$, the two-point correlation function becomes
\begin{align}
    \langle h^\pm (\tau_0, {\mathbf{k_1} }) h^\pm (\tau_0, {-\mathbf{k_1}})\rangle' &= \frac{ e^{2\pi\xi}}{32\pi^2k^3} \left(\frac{H}{M_{\rm Pl}}\right)^4 \int_{q=0}^{\infty} dq\ q  I_{\theta}^\pm(q) \left|  I_{x}(\theta,q) \right|^2, \label{2PCFtensor}
\end{align}
where the integrals with respect to $\theta$ and $x$ are defined as,
\begin{align}
    I_{\theta}^\pm(q) &\equiv \int_{\theta=0}^{\pi} d \theta \frac{\sin\theta(1\pm\cos\theta)^2}{\sqrt{1+q^2-2q\cos\theta}} \left( 1\pm\frac{1-q\cos\theta}{\sqrt{1+q^2-2q\cos\theta}}  \right)^2, \\
    I_{x}(\theta,q) &\equiv\int_{x=0}^{\infty} dx (\sin{x}-x\cos{x})\ \partial_x W_{-i\xi, i\mu}(-2iqx)\ \partial_x W_{-i\xi, i\mu}(-2i\sqrt{1+q^2-2q\cos\theta}x).
\end{align}

\subsubsection{Three-point correlation function}
Using the dominant real mode function approximation, the three-point function can be expressed as
\begin{align}
    \left\langle h^\lambda (\tau_0, \mathbf{k_1}) h^\lambda (\tau_0, \mathbf{k_2}) h^\lambda (\tau_0, \mathbf{k_3})\right\rangle' &= \frac{64H^6}{M_{\rm Pl}^6(2\pi)^{9}k^6y_2^3y_3^3} \nonumber \\
    &\times \int d^3p
    \left[\boldsymbol{\epsilon}^-(\hat{k}_1)\cdot\boldsymbol{\epsilon}^+(\hat{k}_1-\mathbf{p})\right]\left[\boldsymbol{\epsilon}^-(\hat{k}_1)\cdot\boldsymbol{\epsilon}^+(\mathbf{p})\right]\nonumber \\
    &\times \left[\boldsymbol{\epsilon}^-(\mathbf{k_2})\cdot\boldsymbol{\epsilon}^+(-\hat{k}_1+\mathbf{p})\right]\left[\boldsymbol{\epsilon}^-(\mathbf{k_2})\cdot\boldsymbol{\epsilon}^+(-y_3\hat{k}_3-\mathbf{p})\right]\nonumber\\
    &\times \left[\boldsymbol{\epsilon}^-(\mathbf{k_3})\cdot\boldsymbol{\epsilon}^+(y_3\hat{k}_3+\mathbf{p})\right]\left[\boldsymbol{\epsilon}^-(\mathbf{k_3})\cdot\boldsymbol{\epsilon}^+(-\mathbf{p})\right]\ I_{x_1}\ I_{x_2}\ I_{x_3},
\end{align}
where the three integrals are defined as
\begin{align}
   I_{x_1} &\equiv \int dx_1 (\sin{x_1}-x_1 \cos{x_1}) \partial_{x_1} A_+\left(-2i|\hat{k}_1-\mathbf{p}|x_1 \right) \partial_{x_1} A_+(-2i p x_1 ), \\ I_{x_2} &\equiv \int dx_2 (\sin{x_2}-x_2 \cos{x_2}) \partial_{x_2} A_+^*\left(-2i\frac{|\hat{k}_1-\mathbf{p}|}{y_2}x_2 \right) \partial_{x_2} A_+\left(-2i \frac{|y_3\hat{k}_3+\mathbf{p}|}{y_2} x_2 \right), \\
   I_{x_3} &\equiv \int dx_3 (\sin{x_3}-x_3\cos{x_3}) \partial_{x_3} A_+^*\left(-2i\frac{|y_3\hat{k}_3+\mathbf{p}|}{y_3}x_3 \right) \partial_{x_3} A_+^*\left(-2i \frac{|\hat{k}_1-\mathbf{p}|}{y_3} x_3\right).
\end{align}

\subsubsection{Squeezed limit}
Similar to the scalar case, we set the momentum as in \cref{k12sq}, \eqref{k3sq} and \eqref{psq}, and use the approximation in \cref{sqcond}. Then the three point correlation function becomes
\begin{align}
    \langle h^\lambda (\tau_0, \mathbf{k_1}) h^\lambda (\tau_0, \mathbf{k_2}) h^\lambda (\tau_0, \mathbf{k_3})\rangle' &= \frac{81\pi H^6}{8M_{\rm Pl}^6(2\pi)^{9}k_1^6}  \nonumber \\ &\times \int dx_1 \left(\sin{x_1}-x_1\cos{x_1}\right) \partial_{x_1}A_+(\tau_1, 1) \partial_{x_1}A_+(\tau_1,y_3)\nonumber \\ 
    &\times \int dx_2 \left(\sin{x_2}-x_2\cos{x_2}\right) \partial_{x_2}A_+^*(\tau_2, 1) \partial_{x_2}A_+(\tau_2,y_3) \nonumber \\
    &\times \int dx_3 \left(\sin{x_3}-x_3\cos{x_3}\right) \partial_{x_3}A_+^*(\tau_3, y_3) \partial_{x_3}A_+^*(\tau_3,y_3). 
\end{align}
We have simplified the angular part (polarization vectors) and it contributes a factor of $-{27\pi}/{512}$. All the integrals are independent of each other and can be done separately. This greatly decreases the amount of computing power required to calculate the three point function in the squeezed limit.

\subsection{Numerical Computation of the Correlation Functions}
Since the mode function of the massive gauge boson is given in terms of the Whittaker W function, the final expressions for the two- and three-point correlation function also contains these functions inside the integrals with respect to conformal time and loop momenta. For massive fields, it is difficult to approximate the Whittaker functions in terms of closed-form analytic functions, and the integrals are performed numerically. Here we discuss some subtleties with the numerical computation.

In all $n$-point correlation functions for both scalar and tensor modes, we encounter integrals with respect to $x_i \equiv -k_i\tau_i$ and loop momentum $p$, which range from $0$ to $\infty$. For numerical computation, we must set a finite upper limit to these integrals. 

For the integral with respect to $x$, we are only interested in the region where the mode function experiences tachyonic instability. Following the discussion in \cref{sec:2}, we set the upper limit of $x$ to be 
\begin{align}
    x_{\rm max} = \xi + \sqrt{\xi^2 - \left(\frac{m_A}{H} \right)^2}, \label{intlim}
\end{align}
which, in the massless limit, becomes $2\xi$ \cite{Barnaby:2011vw}. This captures the essential physics of gauge field production during inflation and avoids the region dominated by vacuum modes.

For the loop momentum integral, we observe that the integrand with respect to $p$ is a concave functions and falls off rapidly for $p \gg 1$. We have therefore set a hard cut-off $p=20$ and verified that increasing  the cut-off has negligible impact on the result. 

Furthermore, the integrands involve complex Whittaker W functions, and/or their derivatives. We emphasize that using the late-time approximation for the Whittaker W function, as done sometimes in literature, for the computation of $n$-point correlation functions is problematic in general. The late time limit of this function is oscillatory with respect to $x \equiv -k\tau$ with an \emph{increasing} envelope, and near $x = 1$ where particle production happens, it can overestimate the mode function. We use the $\texttt{WhittakerW}$ function available in $\texttt{Mathematica}$ version $12.1$ for our numerical computation.

Finally, the computation of $n$-point correlation functions is, in general, time consuming. To speed-up the process, we parallelize the computation utilizing a cluster computer \cite{HPC}.
 

\section{Generalized Starobinsky Model parameters}\label{model_parameters}
We will closely follow Ref.~\cite{Domcke:2016bkh} to derive the parameters $\gamma$ and $V_0$ of the generalized Starobinsky potential. For simplicity we set $M_{\rm Pl} = 1$ in this appendix, and implicitly assume that all mass-dimension parameters are in units of $M_{\rm Pl}$.

In terms of the potential, the slow-roll parameter, scalar power spectrum and spectral index are given by
\begin{align}
    \epsilon_V &= \frac{1}{2}\left(\frac{V'(\phi)}{V(\phi)}\right)^2,\\
    P_\zeta &= \frac{1}{24\pi^2}\frac{V(\phi)}{\epsilon_V(\phi)},\label{eq:delta}\\
    n_s &= 1 - \frac{2}{N_*} - 6\epsilon \approx 1 - \frac{2}{N_*},
\end{align}
respectively. Here $N_*$ is the number of efolds left before the end of inflation when CMB modes left the horizon in the absence of gauge fields. For $n_s = 0.96$ \cite{Planck:2018jri}, we get $N_* \approx 50$. 

To compute $\gamma$ we start with the slow-roll parameter $\epsilon_V$. For the Starobinsky potential the lowest order expansion gives us,
\begin{align}
    \epsilon_V = \frac{1}{2\gamma^2N_*^2}.
    \label{eq:epsilonv}
\end{align}
From the CMB limit on the tensor to scalar ratio, $r < 0.056$ \cite{Planck:2018jri}, and using $r \approx 16 \epsilon_V$, which holds under slow roll approximation and when the gauge field contribution to the power spectrum is negligible at the CMB scales, we get
\begin{align}
    \gamma^2 \gtrsim \frac{2}{35}.
\end{align}
For numerical results shown in this paper we explicitly choose $\gamma = 0.3$. To compute $V_0$ we start with \cref{eq:delta} and approximate $V(\phi)\approx V_0$. Using \cref{eq:epsilonv} we get,
\begin{align}
    P_\zeta &\approx \frac{V_0\gamma^2N_*^2}{12\pi^2} = 2.5\times 10^{-9}\\
    \Rightarrow V_0 &\approx \gamma^{-2} \times 10^{-10}
\end{align}
To calculate $\phi_{\rm{CMB}}$ we can use the full expression of $V(\phi)$ now that $V_0$ and $\gamma$ are determined. We do this by setting the expression for $P_\zeta$ given by \cref{eq:delta} to $2.5\times 10^{-9}$ when $\phi = \phi_{\rm{CMB}}$. This yields $\phi_{\rm{CMB}} = -7.6$.

Finally, we can calculate the scale $\Lambda$ of the Chern-Simons coupling   requiring $\epsilon_V \approx \epsilon_\phi$, which yields 
\begin{align}
     \frac{1}{2}\left(\frac{\dot{\phi}_0}{H}\right)^2 &\approx \frac{1}{2\gamma^2N_*^2}.
\end{align}
Using the definition of the chemical potential to replace ${\dot{\phi}_0}/{H}$ we get,
\begin{align}
    {\Lambda} \approx \frac{1}{100\gamma\xi}.
\end{align}

\bibliography{references}
\newpage
\bibliographystyle{JHEP}
\end{document}